\documentclass[useAMS,usenatbib,twocolumns]{mn2e} %
\pdfoutput=0

\usepackage{amsmath,amssymb,graphicx} \usepackage{bm,color}
\usepackage{epstopdf}
\usepackage{ulem}
\usepackage{mathtools}
\usepackage{url}
\usepackage{footnote}
\makesavenoteenv{table}

\graphicspath{{plots/}}

\bibpunct{(}{)}{;}{a}{}{,}

\newcommand{\dd}{{\mathrm d}} 
\newcommand{\average}[1]{\left\langle #1 \right\rangle}
\renewcommand{\vec}[1]{\boldsymbol{#1}} \newcommand{\vt}{\vartheta}
\newcommand{\ve}{\varepsilon} \newcommand{\vp}{\varphi}
\newcommand{\ii}{{\rm i}} \newcommand{\ee}{{\rm e}}
\newcommand{\vvt}{\vec \vartheta}

 \newcommand{\Map}{M_{\rm ap}}
\newcommand{\ap}{\rm ap} 

\newcommand{\Omegam}{\Omega_{\rm m}}

\newcommand{\mat}[1]{\textbfss{#1}}

\title [CFHTLenS: cosmic shear two-point and three-point correlation]
{CFHTLenS: Cosmological constraints from a combination of cosmic shear
  two-point and three-point correlations}

\author[L.~Fu et~al.] 
 {
  \parbox[h]{\textwidth}
  {
    Liping Fu$^1$ \thanks{E-mail: fuliping@shnu.edu.cn; fu.lipingmiu@gmail.com},
    Martin Kilbinger$^{2, 3}$ \thanks{E-mail: martin.kilbinger@cea.fr},
    Thomas Erben$^4$,
    Catherine Heymans$^5$,
    Hendrik Hildebrandt$^{6, 4}$,
    Henk Hoekstra$^{7}$,
    Thomas D.~Kitching$^8$,
    Yannick Mellier$^{3, 2}$,
    Lance Miller$^{9}$,
    Elisabetta Semboloni$^{7}$,
    Patrick Simon$^{4}$,
    Ludovic Van Waerbeke$^6$, 
   Jean Coupon$^{10}$, 
   Joachim Harnois-D\'eraps$^{11, 6}$,
    Michael J. Hudson$^{12, 13}$,
    Konrad Kuijken$^{7}$,
    Barnaby Rowe$^{14}$,
    Tim Schrabback$^{7, 15, 4}$,
    Sanaz Vafaei$^{6}$,
    Malin Velander$^{7, 9}$
}
  \vspace*{10pt} \\
  \hspace{-.1cm}$^1$ Shanghai Key Lab for Astrophysics, Shanghai Normal
           University, 100 Guilin Road, 200234, Shanghai, China \\
  \hspace{-.1cm}$^2$ CEA/Irfu/SAp Saclay, Laboratoire AIM, 91191 Gif-sur-Yvette, France\\
  \hspace{-.1cm}$^3$ Institut d'Astrophysique de Paris, UMR7095 CNRS,
           Universit\'e Pierre \& Marie Curie, 98 bis boulevard Arago, 75014 Paris,
           France \\
  \hspace{-.1cm}$^{4}$ Argelander-Institut f\"ur Astronomie, Universit\"at Bonn,
        Auf dem H{\"u}gel 71, 53121 Bonn, Germany \\
\hspace{-.1cm}$^5$ Scottish Universities Physics Alliance, Institute for Astronomy, University of Edinburgh, Royal Observatory, Blackford Hill, \\
\hspace{.15cm}Edinburgh, EH9 3HJ, UK\\
    \hspace{-.1cm}$^{6}$ University of British Columbia, Department of
           Physics and Astronomy, 6224  Agricultural Road, Vancouver, B.C. V6T
           1Z1, Canada \\
   \hspace{-.1cm}$^{7}$ Leiden Observatory, Leiden University, Niels
           Bohrweg 2, 2333 CA Leiden, The Netherlands \\
   \hspace{-.1cm}$^8$ Mullard Space Science Laboratory, University College London, Holmbury St Mary, Dorking, Surrey RH5 6NT, UK \\
   \hspace{-.1cm}$^{9}$ Dept. of Physics, Oxford University, Keble Road, Oxford OX1 3RH, UK \\
   \hspace{-.1cm}$^{10}$ Astronomical Observatory of the University of Geneva, ch. d’Ecogia 16,
1290 Versoix, Switzerland\\
   \hspace{-.1cm}$^{11}$ Canadian Institute for Theoretical Astrophysics, University of Toronto, M5S 3H8, Ontario, Canada \\
   \hspace{-.1cm}$^{12}$ Department of Physics and Astronomy, University of Waterloo, Waterloo, ON, N2L 3G1, Canada \\
   \hspace{-.1cm}$^{13}$ Perimeter Institute for Theoretical Physics,
   31 Caroline Street N, Waterloo, ON, N2L 1Y5, Canada \\
  \hspace{-.1cm}$^{14}$ Department of Physics and Astronomy, University College London, Gower Street, London WC1E 6BT, UK \\
 \hspace{-.1cm}$^{15}$ Kavli Institute for Particle Astrophysics and Cosmology, Stanford University, 382 Via Pueblo Mall,
 Stanford, CA 94305-4060, USA \\
}

\begin{document}

\date{\today}

\pagerange{\pageref{firstpage}--\pageref{lastpage}} \pubyear{2009}

\maketitle

\label{firstpage}

\begin{abstract}


Higher-order, non-Gaussian aspects of the large-scale structure carry
valuable information on structure formation and cosmology, which is
complementary to second-order statistics. In this work we measure
second- and third-order weak-lensing aperture-mass moments from
CFHTLenS and combine those with CMB anisotropy probes.
The third moment is measured with a significance of $2\sigma$.
The combined constraint on $\Sigma_8 = \sigma_8 (\Omega_{\rm m}/0.27)^\alpha$
is improved by 10\%, in comparison to the second-order only, and the allowed
ranges for $\Omega_{\rm m}$ and $\sigma_8$ are substantially reduced.
Including general triangles of the lensing bispectrum yields tighter
constraints compared to probing mainly equilateral triangles.
Second- and third-order CFHTLenS lensing measurements improve
Planck CMB constraints on $\Omega_{\rm m}$
and $\sigma_8$ by 26\% for flat $\Lambda$CDM. For a model with free curvature, the
joint CFHTLenS-Planck result is $\Omega_{\rm m} = 0.28 \pm 0.02$ (68\% confidence),
which is an improvement of 43\% compared to Planck alone.
We test how our results are potentially subject to three astrophysical sources of
contamination: source-lens clustering, the intrinsic alignment of galaxy
shapes, and baryonic effects. We explore future limitations of the
cosmological use of third-order weak lensing, such as the nonlinear model
and the Gaussianity of the likelihood function.

\end{abstract}


\vspace*{3em}
\section{Introduction}
\label{sec:intro}

The extraordinary rise of observational cosmology over the past twenty
years has profoundly modified the ambitions and the methods of
physical cosmology. It has opened a new era where precision cosmology
may allow astronomers and physicists to address key questions about
fundamental laws of physics. Cosmological surveys probing a range of
different scales and epochs, using techniques such as CMB
anisotropies, supernovae of type Ia, baryonic acoustic oscillations,
galaxy cluster counts, and weak gravitational lensing have based a
broad cosmological paradigm upon strong observational foundations.

Cosmological weak lensing, also called cosmic shear, denotes tiny
shape distortions of distant galaxy images that arise from
gravitational lensing of light by the large-scale structure of the
Universe. It is a cumulative, anisotropic gravitational shear effect
that a light bundle experiences by passing through cosmic structures
on the way from the galaxy to the observer. A circular beam of light
is hereby transformed into a small ellipse. This gives us a powerful
way to indirectly observe dark matter in the Universe and to study its
distribution on cosmological scales.

Although the lensing effect is very weak, it modifies the shapes of
galaxies in a coherent manner and can therefore be detected, analysed
statistically, and interpreted within a cosmological model, by
observing millions of galaxies. The distribution of weak gravitational
distortions as a function of angular scale is indeed an almost direct
gravitational imprint of the dark matter distribution projected on the
sky. The second-order shear correlation between galaxy pairs have been
measured from different surveys since 2000, and have been successfully
used to constrain the power spectrum of dark matter.  Recent results
are for example: \cite{FSHK08} for the CFHTLS third data release;
\cite{SHJKS09} for the Hubbble Space Telescope
COSMOS\footnote{\url{http://cosmos.astro.caltech.edu}} survey;
\cite{CFHTLenS-2pt-tomo} and \cite{CFHTLenS-2pt-notomo} for CFHTLenS;
and \cite{2012arXiv1210.2732J} for the Deep Lens Survey.

Third-order cosmic shear statistics contain information about the
bispectrum of the projected matter density, which is the lowest-order
measure of non-Gaussianity of the large-scale structure
\citep{1997A&A...322....1B, 1999A&A...342...15V, 2001MNRAS.322..918V}.
The accuracy of cosmological parameters constraints from combined
measurements of second- and third-order shear statistics is expected
to be increased significantly \citep{2004MNRAS.348..897T, KS05,
  2010APh....32..340V}.

The first detections of third-order shear statistics was obtained from
VIRMOS \citep{2003A&A...397..405B, 2003ApJ...592..664P} and CTIO data
\citep[][hereafter JBJ04]{JBJ04}. With the improvement of shape
measurement techniques and point spread function corrections for
space-based observations by \cite{SHJKS09}, \cite{2011MNRAS.410..143S}
obtained cosmological constraints from three-point shear statistics
using the data from COSMOS, which are consistent with the WMAP7
best-fit cosmology. \cite{CFHTLenS-kappa-maps} measured third-,
fourth- and fifth-order cosmic shear statistics from reconstructed
convergence maps, and found good agreement for the third-order moment
with WMAP7 predictions.

In this paper, we perform a combined second- and third-order
weak-lensing analysis to constrain parameters of different
cosmological models using the CFHT Lensing
Survey\footnote{\url{www.cfhtlens.org}} (CFHTLenS), which covers 154
square degrees in five optical bands $u^\star, g', r', i', z'$
obtained as part of the CFHT Legacy Survey.  A companion paper,
\cite{CFHTLenS-3pt}, presents in more detail the third-moment
measurement and systematics tests.  An overview of the CFHTLenS data
and analysis can be found in \cite{CFHTLenS-data} and
\cite{CFHTLenS-sys}.

This paper is organised as follows.  In Sect.~2, we briefly review the
theoretical background of weak gravitational lensing and
second-/third-order statistics of cosmic shear. In Sect.~3 we describe
the CFHTLenS data and covariance measurement methods and calibration,
the theoretical model, and the statistical analysis to compare our
models to the data.  Section 4 presents the CFHTLenS measurements and
cosmological constraints. In Sect.~5 we discuss astrophysical
contaminants to third-order lensing, and Sect.~6 shows combined
constraints with other cosmological probes. We conclude the paper with
a discussion of our results in Sect.~7.

The data that are presented in this work (aperture-mass moments and covariance matrices) are
available at \texttt{http://www.cfhtlens.org}. The software used for the cosmological
analysis can be downloaded from \texttt{http://cosmopmc.info}.

\section{Weak cosmological lensing}
\label{sec:cs}

\subsection{Theoretical predictions}
\label{sec:cs_theory}

Cosmic shear is the weak lensing effect caused by the large-scale
structure.  The theory of weak lensing has been reviewed in detail in
\cite{BS01,2008ARNPS..58...99H,2008PhR...462...67M,2010CQGra..27w3001B}.

The convergence of a galaxy at angular position $\vec \vartheta$ and
comoving distance $w$ is given by the all matter density contrast
$\delta$ times the lensing efficiency, integrated over all (lens)
distances,
\begin{align}
  \kappa(\vec \vt, w) & = \frac 3 2 \Omegam \frac{H_0} c \int_0^w
  \dd w^\prime g(w^\prime, w)
  \delta(f_K(w^\prime) \vec \vt, w^\prime);
  \label{kappa_w} \\
  g(w^\prime, w) & = \frac{H_0}{c}
  \frac{f_K(w^\prime) f_K(w - w^\prime)}{f_K(w) a(w^\prime)},
  \label{g}
\end{align}
where $f_K(w)$ is the comoving angular diameter distance which depends
on the curvature $K$ of the Universe. $H_0$ is the Hubble constant,
$c$ the speed of light, $\Omegam$ the total matter density, and $a(w)$
the scale factor.  The convergence of a population of sources with a
random density distribution in comoving coordinates $p(w) \dd w$ is
\begin{equation}
  \kappa(\vec \vt) = \int_0^{w_{\rm lim}} \dd w \, p(w) \kappa(\vec
  \vt, w),
  \label{kappa}
\end{equation}
where $w_{\rm lim}$ is the limiting distance of the survey.

The power spectrum $P_\kappa$ of the convergence (\ref{kappa}) is given as
\begin{equation}
  \left\langle 
 \hat \kappa(\vec s) \hat \kappa(\vec s^\prime) \right\rangle
  = \left( 2 \pi \right)^2 \delta_{\rm D}(\vec s + \vec s^\prime) P_\kappa(s),
  \label{kk}
\end{equation}
where $s$ is the modulus of a two-dimensional wave vector
perpendicular to the line of sight.  $P_\kappa$ can be written as a
projection of the power spectrum of dark matter $P_\delta$ along the
line of sight, using the approximation of Limber's equation
\citep{1992ApJ...388..272K}, as defined in \cite{1998MNRAS.296..873S}
\begin{equation}
  P_\kappa(s) =
  \frac{9 \, \Omegam^2 H^4_0}{4 \, c^4}
  \int_0^{w_{\rm lim}} \dd w \, \frac{G^2(w)}{a^2(w)} \,
  P_\delta\left(k = \frac{s}{f_K(w)}; w\right).
  \label{eq:Pk}
\end{equation}
Here, $G(w)$ is the lens efficiency,
\begin{equation}
  G(w) =  \int_w^{w_{\rm lim}} \dd w^\prime p(w^\prime)
  \frac{f_K(w^\prime - w)}{f_K(w^\prime)},
\end{equation}

The bispectrum $B_\kappa$ of the convergence is defined by the
following equation:
\begin{align}
  \MoveEqLeft
   \average{\hat \kappa(\vec s_1) \hat \kappa(\vec s_2) \hat \kappa(\vec
      s_3)}  = (2 \pi)^2 \delta_{\rm D}(\vec s_1 + \vec s_2 + \vec s_3)
  \nonumber \\
  & \times
   \left[ B_\kappa(\vec s_1, \vec s_2) +
    B_\kappa(\vec s_2, \vec s_3) + B_\kappa(\vec s_3, \vec s_1)
  \right].
  \label{kkk}
\end{align}
Using again Limber's equation, $B_\kappa$ is related to the matter
bispectrum $B_\delta$.

To model the highly non-linear bispectrum $B_\delta$ on small scales,
we employ the hyper-extended perturbation theory
\cite[HEPT]{{2001MNRAS.325.1312S}}.  This framework provides functions
to interpolate between the linear regime, where tree-level
perturbation theory is a good description of the bispectrum, and the
strongly non-linear regime. HEPT on these very small scales falls back
on the stable clustering hypothesis, where clustering is assumed to
have reached virialized equilibrium \citep{pee80}.

The original HEPT bispectrum is based on the non-linear power spectrum
fitting formulae from \cite{pd96}. Although the HEPT bispectrum is
expressed as a function of the non-linear power spectrum, the HEPT
coefficients have been fitted to the reduced bispectrum, minimising
the dependence on the power spectrum. Therefore, different
prescriptions for the non-linear power spectrum can be combined with
HEPT, for example the widely-used halofit
\citep{2003MNRAS.341.1311S}. Recently, \cite{2013PhRvD..87l3538S} have
shown that HEPT provides a much better fit to the convergence
bispectrum when using the revised \texttt{halofit} version of
\cite{2012ApJ...761..152T}. These revised fitting functions also match
more closely the convergence power spectrum.  Whereas the original
halofit prescription underestimates power on small scales
\cite[e.g.][]{2012ApJ...761..152T}, the revised halofit overestimates
it slightly \citep{2014ApJ...780..111H}.  An alternative prescription
of the non-linear power spectrum is given by the Coyote emulator
\citep{2014ApJ...780..111H}. Further, a revised version of HEPT was
recently published by \citet{2012JCAP...02..047G}. In
App.~\ref{sec:scales_clone} we test those different models for
$P_\delta$ and $B_\delta$ on N-body simulations. We choose the
combination models of \cite{2012ApJ...761..152T} and
\cite{2014ApJ...780..111H} since it provides the best match to the
CFHTLenS Clone simulations \citep{CFHTLenS-Clone}.  We do not consider
the effect of baryons on the power- and bispectrum.  Their influence
on the matter clustering is important, in particular on small
scales. This behavious can be modeled using hydro-dynamical $N$-body
simulations \citep{2011MNRAS.417.2020S}. Their potential influence is
estimated in Sect.~\ref{sec:baryons}.

\subsection{Second- and third-order functions}
\label{sec:2+3pt}

\subsubsection{Correlation functions}
\label{sec:corr-estim}

The basic observables from a weak-lensing galaxy survey are the
ellipticities $\vec \varepsilon_i$ at galaxy positions $\vvt_i$. From
that it is possible to create a map of the convergence $\hat \kappa$
in Fourier space and to measure the power spectrum and bispectrum by
taking moments.  Such a convergence reconstruction has been performed
recently using CFHTLenS data, and moments of the convergence up to
order 5 have been measured \citep{CFHTLenS-kappa-maps}.

We choose a different approach, which does not require the treatment
of masks and smoothing of the shear field. From the galaxy
ellipticities, we directly estimate the shear second- and third-order
correlation functions, $\xi_\pm$ and $\Gamma^{(0,1,2,3)}$,
respectively.  For the second-order case (2PCFs), we update the
results from \citet[][hereafter K13]{CFHTLenS-2pt-notomo}, using 120
instead of 129 fields, which are the fields that pass the systematics
test on both second- and third-order \citep{CFHTLenS-3pt}.  The
third-order correlation functions (3PCF) are given for a triangle, and
have eight components \citep{tpcf1, 2003ApJ...584..559Z,
  2003ApJ...583L..49T}. We use the four complex {\it natural
  components} as introduced in \cite{tpcf1}. Following the notation of
JBJ04, for triangle vertices $\vec X_1, \vec X_2, \vec X_3$, we define
two triangle side vectors as $\vec s = \vec X_2 - \vec X_1$, $\vec t =
\vec X_3 - \vec X_2$.  An unbiased estimator for the zero-th component
is
\begin{equation}
 \hat \Gamma^{(0)}(\vec s, \vec t) = \frac{ \sum_{ijk} w_i \, w_j \, w_k \,
      \ve_i \, \ve_j \, \ve_k \, {\rm e}^{-6 {\rm i} \alpha }}
					 { \sum_{ijk} w_i \, w_j \, w_k},
  \label{gamma0}
\end{equation}
 where $w$ is the weight of shear of each galaxy.  As in JBJ04, we
 choose the polar angle $\alpha$ of the triangle side $\vec s$ to be
 the projection angle for all vertices. The sum is performed over
 triples of galaxies $i, j, k$ which form triangles that are close to
 $(\vec s, \vec t)$ within the chosen binning scheme. We use the
 tree-based code, kindly provided by M.~Jarvis, to perform this
 summation. The binning scheme is detailed in JBJ04, see also
 App.~\ref{sec:binning}.

The first component is estimated as
\begin{equation}
 \hat \Gamma^{(1)}(\vec s, \vec t) = \frac{ \sum_{ijk} w_i \, w_j \, w_k
      \ve_i^\ast \, \ve_j \, \ve_k \, {\rm e}^{-2{\rm i} \alpha }}
					 { \sum_{ijk} w_i \, w_j \, w_k}.
  \label{gamma1}
\end{equation}
The other two components $\hat \Gamma^{(2,3)}$ are obtained from $\hat
\Gamma^{(1)}$ by cyclic permutations of the triangle parameters \citep{tpcf1}.

\subsubsection{Aperture moments}

The aperture mass, introduced by \cite{KSFW94} and
\cite{1996MNRAS.283..837S}, is a scalar quantity expressed in terms of
convergence $\kappa$ inside an aperture centred at some point $\vvt$,
filtered by a function $U_\theta$ that depends on some characteristic
smoothing scale $\theta$. If $U_\theta$ is compensated, i.e.~$\int \dd
\vt \, \vt \, U_\theta(\vt) = 0,$ the aperture mass can be expressed
in terms of the tangential shear component, $\gamma_{\rm t}(\vec
\vt^\prime) = - \Re [ \gamma(\vec \vt^\prime) \exp(-2\ii\vp) ]$, where
$\vp$ is the polar angle of the vector $\vec \vt^\prime - \vec \vt$,
\begin{align}
   \Map(\theta, \vvt) = & \int \dd^2 \vt^\prime \, U_\theta(|\vvt-\vvt^\prime|) \, \kappa(\vec \vt^\prime)
   \nonumber \\
   = 
   &
   \int \dd^2 \vt^\prime \, Q_\theta(|\vvt-\vvt^\prime|) \, \gamma_{\rm t}(\vec \vt^\prime).
   \label{map_defx}
\end{align}
The filter function $Q_\theta$ is given in terms of $U_\theta$, see
\cite{KSFW94} and \cite{1996MNRAS.283..837S}.  Correspondingly,
$M_\times$ is defined in terms of the cross-component of the shear,
$\gamma_\times(\vec \vt)=-\Im [ \gamma(\vec \vt^\prime) \exp(-2\ii\vp)
]$, and is a measure of the B-mode,
\begin{equation}
 M_\times(\theta, \vvt)  = \int \dd^2 \vt^\prime \, Q_\theta(|\vvt-\vvt^\prime|) \,
 \gamma_\times(\vec \vt^\prime). 
  \label{eq:mapB_def}
\end{equation}
The aperture-mass dispersion can be calculated in terms of the
covergence power specturm $P_\kappa$,
\begin{equation}
\left\langle{M_{\rm ap}^2}\right\rangle(\theta) = \int \frac{\dd \ell \, \ell}{2\pi}
P_\kappa(\ell) \hat U^2(\theta \ell).
\label{map2-power}
\end{equation}
where $\hat U$ is the Fourier transform of $U_\theta$.  The
third-order moment of the aperture-mass has been introduced by JBJ04
and \cite{2003ApJ...592..664P}. Its generalisation involves the
correlation of the aperture-mass for three different smoothing scales,
which optimally probes the bispectrum for general triangles, has been
defined in \citet[][hereafter SKL05]{SKL05}. It can be written as
\begin{align}
  \MoveEqLeft
   \left\langle{M_{\rm ap}^3}\right\rangle(\theta_1, \theta_2, \theta_3) \equiv
   \average{M_{\rm ap}(\theta_1)M_{\rm ap}(\theta_2)M_{\rm ap}(\theta_3)}
   \nonumber \\
   = & \int \frac{\dd^2\ell_1}{(2\pi)^2}
   \int \frac{\dd^2\ell_2}{(2\pi)^2}\,B_\kappa(\vec\ell_1,\vec\ell_2)
   \nonumber \\
   &
   \times
   \!\!
   \sum\limits_{(i,j,k) \in S_3}
   \!\!
   \hat
   U(\theta_i|\vec\ell_1|)\,\hat U(\theta_j|\vec\ell_2|)\,
   \hat U(\theta_k |\vec\ell_1+\vec\ell_2|),
   \label{map3-bi}
\end{align}
where $S_3$ is the symmetric permutation group of $(123)$.  One of the
four integrals in Eq.~(\ref{map3-bi}) can be performed analytically
using the angular dependence of the bispectrum due to the statistical
isotropy of the convergence field. The result is given in
\cite{KS05}. The simplest expressions for the third-order aperture-mass moment in
terms of $B_\kappa$ exist for a Gaussian-shaped filter function
$U_\theta$,
\begin{equation}
	U_\theta(\vt) = \frac{1}{2\pi\theta^2} \left( 1 - \frac{\vt^2}{2 \theta^2} \right)
			\exp\left(-\frac{\vt^2}{2 \theta^2}\right).
	\label{UG}
\end{equation}

There are several advantages of using aperture moments instead of
$n$-point correlation functions. Most importantly, aperture
measures are sensitive only to the E-mode of the shear field. They
filter out long-wavelength modes where an E-/B-mode separation is not
possible given the finite survey volume \citep{COSEBIs}. They are
therefore less susceptible to some type of systematics in the
data. Further, a theoretical prediction from the convergence
bispectrum $B_\kappa$ is much easier and faster obtained for the
aperture third moment than for the three-point correlation function
(SKL05). It is therefore more efficient to use in a Monte-Carlo
sampling analysis.

\subsection{Measurement of aperture moments}
\label{sec:2+3pt_measure}

The direct measurement of the aperture-mass second and third moments
by averaging over positions $\vec \vartheta$ is not straightforward.
Masks and gaps in the data can cause biases in the estimation, or make
a lot of the area unused. Instead, these moments can be expressed as
integrals over the two- and three-point correlation functions, for
which unbiased estimators have been introduced in
Sect.~\ref{sec:corr-estim}.

For second order, this relation was found by
\cite{2002ApJ...568...20C} and \cite{2002A&A...389..729S},
\begin{equation}
  \langle M_{\rm ap, \times}^2\rangle(\theta) = \frac 1 2 \sum_i \vt_i \, \Delta \vt_i
  \left[ T_+\!\left( {\vt_i} \right) \hat \xi_+(\vt_i) \pm
    T_-\!\left( {\vt_i} \right) \hat \xi_-(\vt_i) \right] ,
  \label{map_EB}
\end{equation}
with the functions $T_{\pm}(x) = \int_0^\infty \dd t \, {\rm J}_{0, 4}(x t) \, \hat U^2(t).$
Analytical expressions corresponding to the Gaussian filter (\ref{UG}) can
be found in \cite{2002ApJ...568...20C},
\cite{2002A&A...389..729S}, and \cite{2003ApJ...592..664P}.

Corresponding relations for the third-order aperture-mass moment have
been derived in JBJ04, and, for the generalised case, in SKL05. 
  First, we define the complex quantity $M(\theta) = M_{\rm
    ap}(\theta) + {\rm i} M_\perp(\theta)$. Next, third moments of $M$
  are calculated as integrals over the 3PCF; from these moments, the
  the E- and B-modes are formed as linear combinations (see
  below). The integrals are performed over all triangle configurations
  $(\vec s, t)$,
\begin{equation}
   \langle M^3 \rangle(\theta_{123}) 
      = S \int \frac{s \dd s}{\Theta^2} \int \frac{\dd^2 t}{\Theta^2} \ee^{-Z} \,
      \Gamma^{(0)}(\vec q_{123}) \, T^0_{123}(\vec s, t),
   \label{mmm-1}
\end{equation}
and 
\begin{equation}
   \langle M^2 M^\ast \rangle(\theta_{123}) 
      = S \int \frac{s \dd s}{\Theta^2} \int \frac{\dd^2 t}{\Theta^2} \ee^{-Z} \,
      \Gamma^{(1)}(\vec s, t) \, T^1_{123}(\vec s, t) .
   \label{mmmc-1}
\end{equation}
Here we have introduced the short forms $\theta_{123} \equiv
(\theta_1, \theta_2, \theta_3)$,  $\vec q_{123} \equiv (\vec q_1,
  \vec q_2, \vec q_3)$, and defined $T^i_{123}(\vec s, t) = T^i(\vec
s, t, \theta_{123})$.  For mathematical convenience we write the
  2D vectors $\vec q_1, \vec s$ and $\vec t$ complex quantities, with
  their real (imaginary) part being the $x$- ($y$-)component. The
triangle orientation in the integrand is chosen such that $\vec t = t
+ 0{\rm i}$. The filter functions $T^i$ can be inferred from
\cite{SKL05}, and are given as
\begin{align}
   T^0_{123}(\vec s, t) = &
       - \frac 1 {24} \frac{\vec q_1^{\ast 2} \vec q_2^{\ast 2} \vec q_3^{\ast 2}}{\Theta^6}
         f_1^{\ast 2} f_2^{\ast 2} f_3^{\ast 2}; \\
   T^1_{123}(\vec s, t) = &
       - \frac 1 {24} \frac{\vec q_1^2 \vec q_2^{\ast 2} \vec q_3^{\ast 2}}{\Theta^6}
         f_1^2 f_2^{\ast 2} f_3^{\ast 2}
      + \frac 1 9 \frac{q_1^2 \vec q_2^\ast \vec q_3^\ast}{\Theta^4}
         f_1 f_2^\ast f_3^\ast g_1^\ast
         \nonumber \\
      &
      - \frac 1 {27} \left(\frac{q_1^{\ast 2} g_1^{\ast 2}}{\Theta^2}
         + \frac{2 \theta_2^2 \theta_3^2}{\Theta^4} \frac{q_2^\ast q_3^\ast}{\Theta^2} f_2^\ast f_3^\ast
         \right),
\end{align}
with
\begin{align}
   f_i = & \; \frac{\theta_j^2 + \theta_k^2}{2\Theta} \frac{(\vec q_j - \vec q_k)\vec q_i}{q_i}
         \frac{\theta_j^2 - \theta_k^2}{6\Theta^2};
         \\
   \quad
   g_i =
   &
   \; \frac{\theta_j^2 \theta_k^2}{\Theta^4} - \frac{(\vec q_j - \vec q_k)\vec q_i^\ast}{q_i}
         \frac{\theta_i (\theta_j^2 - \theta_k^2)}{3\Theta^4}.
\end{align}
The vectors $\vec q_i$ connect the vectices $\vec X_i$ to the triangle
centroid, which are the same vectors as in JBJ04. Further,
\begin{align}
   \Theta = & \left( \frac{\theta_1^2 \theta_2^2 + \theta_2^2 \theta_3^2 + \theta_3^2 \theta_1^2}{3}\right)^{1/4};
   \nonumber \\
   S = & \; \frac{\theta_1^2 \theta_2^2 \theta_3^2}{\Theta^6}; \;\; 
   \nonumber \\
   Z = & \left({6 \Theta^4}\right)^{-1} \left[
     (-\theta_1^2 + 2 \theta_2^2 + 2 \theta_3^2) q_1^2
     \right.
     \nonumber \\
     &
     \left.
     + (2 \theta_1^2 - \theta_2^2 + 2 \theta_3^2) q_2^2
               + (2 \theta_1^2 + 2 \theta_2^2 - \theta_3^2) q_3^2
               \right]
               . \nonumber
\end{align}

As described in JBJ04, the 3PCF is only calculated for one of the six
possible permutations of triangle sides $(\vec s, t)$, given by $s < t
< |t - \vec s|$.  To cover the full range of triangles,
eqs.~(\ref{mmm-1}) and (\ref{mmmc-1}) have to be split up into six
terms, by permuting the centroid vectors $\vec q_i$, see eq.~(59) in
JBJ04. In our case of the generalised third moment, this implies
permuting the smoothing angles $\theta_i$. The result is
\begin{equation}
   \langle M^3 \rangle(\theta_{123}) 
      = 6 S \int \frac{s \dd s}{\Theta^2} \!\!\!\! \int\limits_{s < t
< |t - \vec s|} \!\!\!\! \frac{\dd^2 t}{\Theta^2} \ee^{-Z} \,
      \Gamma^{(0)}(\vec s, t) \, T^0_{(123)}(\vec s, t),
   \label{mmm}
\end{equation}
where the brackets around the indices denote permutations, i.e.~
\begin{equation}
A_{(123)} = \frac{1}{3!} \left( A_{123} + A_{213} + A_{312} + A_{132} + A_{231} + A_{321} \right).
\end{equation}
In eq.~(\ref{mmmc-1}), the permutations of the triangle sides result
in a change of the complex conjugated vertex. The result is
\begin{align}
   \langle M^2 M^\ast \rangle
      = 
      2 S \int \frac{s \dd s}{\Theta^2} \!\!\!\! \int\limits_{s < t
< |t - \vec s|} \!\!\!\! \frac{\dd^2 t}{\Theta^2} \ee^{-Z} 
      \sum_{i=1}^3  \Gamma^{(i)}(\vec s, t) \, T^i_{3(12)}(\vec s, t),
   \label{mmmc}
\end{align}
which is symmetric under permutation of $\theta_1$ and $\theta_2$. For brevity, we omitted the argument
$(\theta_{123})$. Likewise, we have
\begin{align}
   \langle M M^\ast M \rangle
      =      2 S \int \frac{s \dd s}{\Theta^2} \!\!\!\!\! \int\limits_{s < t
< |t - \vec s|} \!\!\!\!\! \frac{\dd^2 t}{\Theta^2} \ee^{-Z}
      \sum_{i=1}^3  \Gamma^{(i)}(\vec s, t) \, T^i_{2(13)}(\vec s, t),
   \label{mmcm}
\end{align}
and
\begin{align}
   \langle M^\ast M^2  \rangle
      =       2 S \int \frac{s \dd s}{\Theta^2} \!\!\!\! \int\limits_{s < t
< |t - \vec s|} \!\!\!\! \frac{\dd^2 t}{\Theta^2} \ee^{-Z} 
      \sum_{i=1}^3  \Gamma^{(i)}(\vec s, t) \, T^i_{1(23)}(\vec s, t).
   \label{mcmm}
\end{align}
In the previous two equations, only the last two indices of the filter
functions $T^i$ are permuted, i.e.~$T^i_{j(kl)} = \frac 1 2 [ T_{jkl}
+ T_{jlk}]$.

As in JBJ04 and SKL05, we combine eq.~(\ref{mmm-1}) -- (\ref{mcmm}) to
obtain the E- and B-mode components of the third-order aperture-mass
moment. The pure E- and B-modes are, respectively,
\begin{align}
  \MoveEqLeft
  EEE: \quad \langle M_{\rm ap}^3\rangle (\theta_{123})
  = \frac 1 4 {\cal R} \Big[
   \langle M^\ast M^2 \rangle
   \nonumber \\
   &
   + \langle M M^\ast M \rangle
   + \langle M^2 M^\ast \rangle
   + \langle M^3 \rangle \Big]
   (\theta_{123}); 
   \label{Map3} \\
   \MoveEqLeft
   BBB: \quad \langle M_\times^3\rangle (\theta_{123})
   = \frac 1 4 {\cal I} \Big[
   \langle M^\ast M^2 \rangle
   \nonumber \\
   &
   + \langle M M^\ast M \rangle
   + \langle M^2 M^\ast \rangle 
   - \langle M^3 \rangle \Big]
   (\theta_{123}). 
	\label{Mx3}
\end{align}
The mixed E-/B-mode components are
\begin{align}
  \MoveEqLeft
  EEB: \quad \langle M_{\rm ap} M_{\rm ap} M_\times\rangle (\theta_{123}) 
	= \frac 1 4 {\cal I} \Big[ \langle M^\ast M^2 \rangle 
      \nonumber \\
      &
      + \langle M M^\ast M \rangle
      - \langle M^2 M^\ast \rangle + \langle M^3 \rangle \Big]
      (\theta_{123}); 
      \label{MapMapMx} \\
  \MoveEqLeft
    EBB: \quad \langle M_{\rm ap} M_\times M_\times\rangle (\theta_{123})
    = \frac 1 4 {\cal R} \Big[ - \langle M^\ast M^2 \rangle
  \nonumber \\
  &
  + \langle M M^\ast M \rangle
  + \langle M^2 M^\ast \rangle - \langle M^3 \rangle \Big]
  (\theta_{123}).
  \label{MapMxMx}
\end{align}
Both mixed components have further permutations, which can be
obtained by permuting the smoothing scales.

The expectation value of the mixed components (\ref{MapMapMx},
\ref{MapMxMx}) is non-zero only if the E-and B-modes are
correlated. For a parity-symmetric shear field, only the last B-mode
component (\ref{MapMxMx}) can be non-zero
\citep{2003A&A...408..829S}. However in practise, noise sample
variance causes a given observed region to violate parity, and all
three B-mode components can be non-zero.

\subsection{E-/B-mode mixing from incomplete coverage of the shear correlation}
\label{sec:mixing}

To estimate the third-order aperture-mass moment from data, we replace
the integrals in eqs.~(\ref{mmm}) -- (\ref{mcmm}) by sums over the
measured triangle configurations. These estimators will be biased
since both on very small and very large scales, triangles cannot be
measured. The former incompleteness occurs on scales of around 10
arcsec, which is the size of the CFHTLenS postage stamps around
galaxies: correlations between objects at separation below this scale
are not measured reliably \citep{CFHTLenS-shapes}. The large-scale
limit is set by the survey size.

For the Gaussian filter (\ref{UG}) and given smoothing angles
$\theta_{123}$, the functions $T^i_{jkl}$ decrease as a Gaussian with
increasing triangle sides. To reduce the bias from incomplete sampling
at large scales, we carry out the integrals to four times the maximal
smoothing angle (JBJ04).

Our smallest smoothing angle is $\theta = 2$ arcmin. This corresponds
to a bias of around 1\% for the aperture-mass dispersion $\langle
M_{\rm ap} \rangle$ using the Gaussian filter \citep{KSE06}.

For third-order, we expect a smaller bias: Firstly, the functions
$T^i_{ijk}$ are proportional to the triangle sides to the sixth power,
compared to $T_{\pm}(x) \propto x^2$, so small scales are more
suppressed. Secondly, very small triangles do not contribute much
because the three-point correlation functions tend to zero for
decreasing triangle size. This is because they filter the bispectrum
with the first-kind Bessel functions ${\rm J}_6$ and ${\rm J}_2$,
respectively, which tend to zero for decreasing angular scales, in
contrast to the case of $\xi_+$ which filters the power spectrum with
$J_{\rm 0}$ approaching unity towards small arguments. A recent
publication shows that the leakage is indeed negligible below
$1\arcmin$ \citep{2014A&A...561A..68S}.

\section{Data and calibration set-up}
\label{sec:data_and_calib}

\subsection{Data}
\label{sec:sub_data}

An overview of the weak-lensing data of the Canada-France-Hawaii
Lensing Survey (CFHTLenS) is given in \cite{CFHTLenS-sys}. See
subsequent papers for details on the data reduction
\citep{CFHTLenS-data}, photometric redshifts \citep{CFHTLenS-photoz},
and galaxy shape measurements \citep{CFHTLenS-shapes}.

CFHTLenS consists of 171 pointings covering 154 square degrees in five
optical bands. For second-order cosmic shear, a sample of 129 fields
was consistent with no remaining systematics \citep{CFHTLenS-sys}.
\cite{CFHTLenS-3pt} analysed systematic contributions to third-order
statistics and found significant systematics in an additional 9 out of
the 129 fields (see their Fig.~5). We therefore choose for our
combined second- and third-order analysis the conservative sample of
120 fields.

As in K13, we select galaxies within the
redshift range $0.2 < z_{\rm p} < 1.3$. This leaves us with 4.2
million source galaxies, corresponding to an effective number density
of $14$ galaxies per square arcmin. For the model of the lensing
signal, the redshift distribution is taken as the sum of the redshift
probability functions over all galaxies, providing a mean redshift of
0.748  (see \citet{CFHTLenS-photoz} for the tests on
  the reliability of the photometric redshifts used in all CFHTLenS
  papers).  

\subsection{Shear calibration}
\label{sec:calib}

We use the calibration of measured galaxy shapes from
\cite{CFHTLenS-sys}, accounting for additive and multiplicative biases
of the estimated ellipticity,
\begin{equation}
  \varepsilon^{\rm obs} = (1 + m) \varepsilon^{\rm true} + c.
\end{equation}
The additive bias for the first component of the ellipticity
$\varepsilon_1$ is found to be consistent with zero, while it shows a
bias at the level of $\sim 2 \times 10^{-3}$ on average, which is
subtracted from the second ellipticity component $\varepsilon_2$ for
each galaxy.
The multiplicative bias $m$ is fitted as a function of the galaxy
signal-to-noise ratio $S/N$ and size $r$, and applied globally
\citep{CFHTLenS-shapes}. The 2PCFs $\xi_\pm$ are corrected for as in
\citet{CFHTLenS-sys} and K13. We calculate the
calibration function
\begin{equation}
  1 + K_2(\theta) = \frac{\sum_{ij} w_i w_j (1 + m_i) (1 + m_j)}
  {\sum_{ij} w_i w_j},
  \label{Kestim}
\end{equation}
where the sum is carried out over pairs of galaxies with separation
within a bin around $\theta$.

Analogeously, the calibration factor $1 + K_3$ for the three point
shear correlation function $\Gamma^{(i)}$ is
\begin{equation}
  1 + K_3(\vec s, t) =
  \frac{\sum_{ijk} w_i w_j w_k (1 + m_i) (1 + m_j)(1 + m_k)}
  {\sum_{ij} w_i w_j w_k},
  \label{Kestime3}
\end{equation}
where the sum goes over all triangles in a bin around $(\vec s, t)$.
We divide all eight components of the 3PCF by $1 + K_3$.  The
multiplicative bias $m$ is on average $-0.08$, so that the 2PCFs and
3PCF are divided by $1 + K_2$ and $1 + K_3$, respectively.  We get
corrections of order $1 + K_2 \sim 0.89$ and $1 + K_3 \sim 0.85$,
virtually independent of angular scale.

\subsection{Data covariance}
\label{sec:cov}

The covariance of the third-order aperture mass contains terms up to
sixth order in the shear. Semi-analytical expressions for those terms
in Fourier space exist using the halo model of dark matter
\citep{2013MNRAS.429..344K, 2013PhRvD..87l3538S}. However, we choose a
numerical approach as follows. We measure aperture-mass moments on
realisations of dark-matter $N$-body and ray-tracing simulations, and
estimate the covariance from the field-to-field scatter. The
simulations are populated by galaxies, with spatial distribution
including masks, redshift distribution, and shape noise corresponding
to observed CFHTLenS characteristics, so that the obtained covariance
matrix includes shape noise and cosmic variance. The N-body and
ray-tracing method that underlies this CFHTLenS ``Clone'' is described
in \cite{CFHTLenS-Clone}. We measure the two-point and three-point
aperture-mass statistics from each of the 184 realisations over the
same angular range as for CFHTLenS data (see
Sect.~\ref{sec:results_third}). To obtain the correlation between
second- and third-order quantities, we measure both statistics
simultaneously and calculate their field-to-field cross-covariance.
The Clone is based on the WMAP5 (+BAO+SN) cosmology. We do not take
into account the variation of the covariance with cosmology
\citep{2009A&A...502..721E}. This effect was found to be minor for
second-order cosmic shear (K13).

The largest available scale from the Clone is 280 arcmin. This
corresponds in principle to a maximum Gaussian smoothing scale for the
aperture-mass of 70 arcmin. However, the signal-to-noise ratio of
$\langle M_{\rm ap}^3 \rangle$ is very small on large scales given the
statistical power of CFHTLenS.  From the Clone we found no significant
improvement when adding scales larger than about 15 arcmin.  For that
reason, we calculate the third-order aperture-mass moment up to only
15 arcmin.

The final covariance matrix is scaled with the ratio of the effective
area $0.9\times 16$ pointings devided by 120 MegaCam pointings which
have passed the systematics test \citep{CFHTLenS-sys}.  This rescaling
is valid strictly only for Gaussian fields, and we are neglecting
couplings between small and large modes.  The relatively small scales
of our data vector should not be affected too much by this.  Further,
to correct for the bias of the inverse covariance estimator
\citep{andersen03,HSS07}, we multiply with the factor $\alpha =
(n-p-2)/(n-1)$, where $n = 184$ is the number of simulated fields, and
$p$ is the number of angular scales. The smallest correction factor,
in the case of the combined data vector with $p = 51$, is $\alpha =
0.72$, which corresponds to a regime where the trace of the de-biased
inverse covariance is accurate to a few percent.  The expected
parameter error uncertainties are less than $15$ percent
\citep{2013MNRAS.tmp.1312T}.

\subsection{Cosmological parameter space}
\label{sec:parameters}

To relate the weak-lensing and external cosmological data to
theoretical models, we use a multi-variate Gaussian likelihood
function,
\begin{align}
\log L(\vec d | \vec p) =
 \left( \vec d - \vec
    y(\vec p) \right)^{\rm t} \mat{C}^{-1}(\vec p) \left( \vec d - \vec
    y(\vec p) \right) + \rm{const},
\label{log_likeli}
\end{align}
where $\vec y(\vec p)$ denotes the theoretical prediction for the data
$\vec d$ for a given $m$-dimensional parameter vector $\vec p$.  For
the CFHTLenS data, $\vec y$ is the vector of measured aperture-mass
second and third moments, or their concatenation, all as function of
angular scales. For the generalised third moment, which is a function
of three smoothing scales $(\theta_i, \theta_j, \theta_k)$, we
construct a vector in lexical order such that $\theta_i \leq \theta_j
\leq \theta_k$. We test and discuss the Gaussian approximation of the
likelihood in App.~\ref{sec:distribution}.

We use \texttt{CosmoPMC}\footnote{\texttt{http://cosmopmc.info}} to
sample the CFHTLenS weak-lensing posterior. For constraints from
CFHTLenS combined with other probes, we importance-sample the WMAP9
and Planck MCMC chain with the CFHTLenS PMC sample. For Planck
\citep{2013arXiv1303.5076P} we use the chain that samples the
combination of Planck temperature data and CMB lensing.

We run the PMC algorithm for up to ten iterations, using 10,000 sample
points in each iteration. To reduce the Monte-Carlo variance, we use
larger samples with 100,000 points for the final iteration. For the
flat $\Lambda$CDM model, the base parameter vector for CFHTLenS weak
lensing is $\vec p = (\Omegam, \sigma_8, \Omega_{\rm b}, n_{\rm s},
h)$. For dark-energy and non-flat models, the parameter vector has one
more parameter $w_0$ and $\Omega_{\rm de}$, respectively. For the
combination with WMAP9 and Planck, the reionisation optical depth
$\tau$ and the Sunyaev-Zel'dovich (SZ) template amplitude $A_{\rm SZ}$
are added to the parameter vector. In this case, we use
$\Delta^2_{\cal R}$ as the primary normalisation parameter, and
calculate $\sigma_8$ as a derived parameter. Moreover, when WMAP9 is
added to CFHTLenS, we use flat priors which cover the high-density
regions and the tails of the posterior distribution well. The priors
of Flat $\Lambda$CDM, Flat $w$CDM and Curved $\Lambda$CDM models are
summarised in Table~\ref{tab:pars}.

 \begin{table}
 \begin{center}
 \caption{The parameters sampled under the weak-lensing CFHTLenS posterior.
   The second column indicates the (flat) prior ranges, for the three
     models analysed in this work (flat $\Lambda$CDM, flat $w$CDM and curved
         $\Lambda$CDM).}
 
\begin{tabular}{@{}lll}
\hline
Param.\           & Prior & Description  \\ \hline
\multicolumn{3}{@{}l}{CFHTLenS, $\Lambda$CDM} \\ \hline
$\Omegam$       & $[0, 1.2]$  & Total matter density \\
$\sigma_8$      & $[0.2, 1.5]$ & Power-spectrum normalisation\\
$\Omega_{\rm b}$ & $[0, 0.1]$  & Baryon density \\
$n_{\rm s}$      & $[0.7, 1.3]$ & Spectral index of prim.\ density fluct.\  \\
$h$             & $[0.4, 1.2]$ & Hubble parameter\\ \hline \\
\multicolumn{3}{l}{Additional parameter for $w$CDM} \\ \hline
$w_0$           &     $[-3.5; 0.5]$ &  Const.\ term in dark-energy eq.\ of state \\ \hline \\
\multicolumn{3}{l}{Additional parameter for curved $\Lambda$CDM} \\ \hline
$\Omega_{\rm de}$ & $[0, 2]$ &  Dark-energy density \\
\hline \\
 \end{tabular}
 \label{tab:pars}
\end{center}

 \end{table}

 We choose and test the angular scale range together with the theoretical model using
the CFHTLenS Clone simulations. More details can be found in App.~\ref{sec:scales_clone}.

\section{CFHTLenS weak-lensing results}
\label{sec:results}

In this section we present the measurement of second- and third-order
aperture-mass measures from CFHTLenS. We show results on the
cosmological parameters $\Omegam$ and $\sigma_8$, the parameters that
are best measured by weak cosmological lensing. We obtain constraints
from second- and third-order statistics, and from their
combination. With current surveys, these two parameters are
near-degenerate, where the direction of degeneracy is approximately a
power law, given by the amplitude parameter $\Sigma_8 = \sigma_8
(\Omega_{\rm m})^\alpha$. We summarise our results on this derived
parameter at the end of Sect.~\ref{sec:contaminants}.

\subsection{Second--order measures}
\label{sec:results_second}

\begin{figure}
  \resizebox{0.9\hsize}{!}{
    \includegraphics[bb = 90 50 380 300]{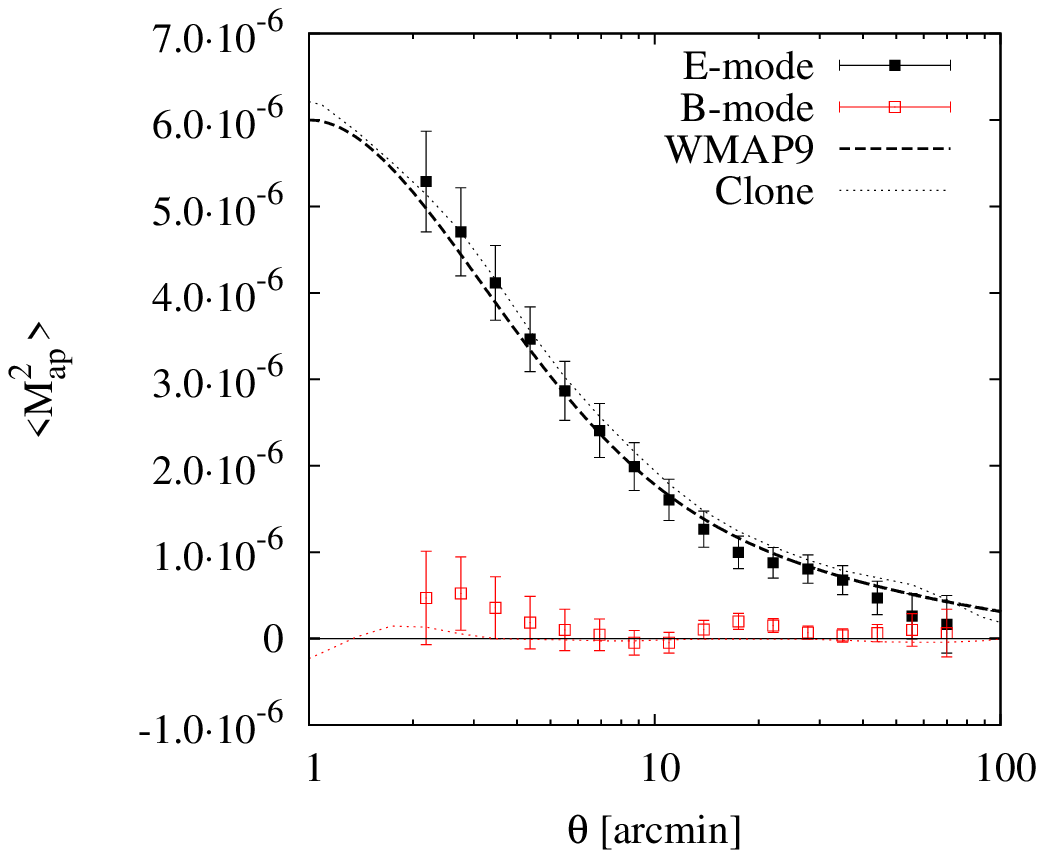}
  }

   \centerline{
  \resizebox{0.99\hsize}{!}{
    \includegraphics[bb = 90 50 380 300]{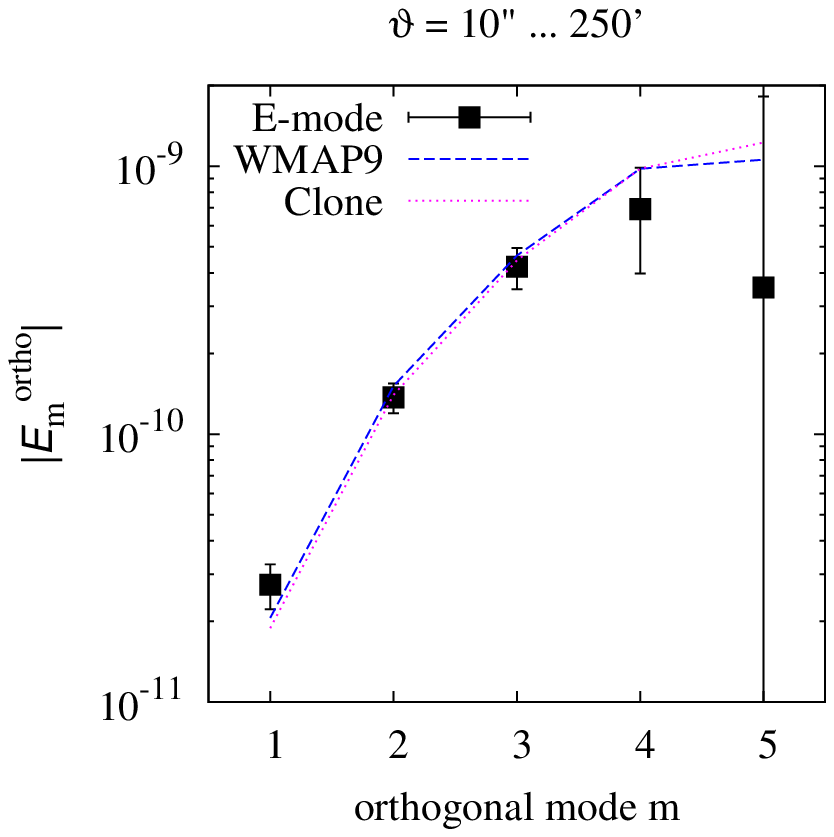}
   }
  }

  \caption{Second-order shear functions measured from CFHTLenS mosaic
    data.  Black filled symbols: E-mode; red open symbols: B-mode.
    The results are compared to the theoretical prediction using the
    WMAP9 cosmology (dashed line), and the E-/B-mode from the Clone
    (dotted lines).  \emph{Upper panel:} The aperture-mass dispersion using
    the Gaussian compensated filter, as a function of smoothing scale
    $\theta$.  \emph{Lower panel:} Orthogonalised COSEBIs (absolute values), ordered by increasing variance.
    as a
    function of orthogonal mode number $m$.  }
   \label{fig:map2:cosebis}
\end{figure}

We report updates of the second-order measurements and resulting
cosmological parameters compared to K13. First, we add a measurement
of the aperture-mass dispersion using the Gaussian filter (\ref{UG}),
which is the filter function we employ for the third-order
aperture-mass moment.  The top panel of Fig.~\ref{fig:map2:cosebis}
shows the E-/B-modes in the angular scale range 2 to 70 arcmin. This
is to combine measurements with the same smoothing function. Note that
there is no necessity for that: for combinations of second- and
third-order measures to obtain the most stringent cosmological
parameter constraints, we use the optimal second-order pure E-mode
measure. These are the so-called COSEBI (Complete Orthogonal Shear E-
and B-mode Integral) modes \citep{COSEBIs} with the logarithmic
filter, for the full available range of angular scales, from 10 arcsec
to 250 arcmin. This measure was presented in K13. The COSEBI modes are
strongly correlated, which makes visual inspection of the data and
comparison to the prediction difficult. Therefore, we show
uncorrelated data points $E^{\rm ortho}_m$ as orthogonal
transformation of the COSEBIs $E_n$, $E^{\rm ortho}_m = S_{mn} E_n$,
where ${\mat S}$ is an orthogonal matrix, ${\mat S} {\mat S}^{\rm T} =
1$.  The result is presented in the lower panel of
Fig.~\ref{fig:map2:cosebis}. Increasing modes $m$ have larger error
bars, which correspond to the elements of the diagonal matrix
${\Sigma}$, obtained by diagonalising the COSEBIs covariance matrix
${\mat C} = {\mat S} {\Sigma} {\mat S}^{\rm T}$.

The covariance matrices of both the aperture-mass dispersion and the COSEBIs
are the field-to-field dispersion from the 184 independent clone simulation
realisations including shape noise, rescaled to the CFHTLenS area.

Second, we update our model of the non-linear power spectrum with the extended
Coyote emulator \citep{2014ApJ...780..111H}, which provides more accurate
estimates of $P(k, z)$ over a wider range in wave number $k$, redshift $z$, and
cosmological parameters, compared to the first version \citep{CoyoteI,
CoyoteII, CoyoteIII}.  We do not include baryonic effects to the power
spectrum. Our smallest scale is $5.5$ arcmin, corresponding to a 3D Fourier
scale $k$ of about $0.7 \, h/$Mpc at redshift of maximum lensing efficiency for
CFHTLenS depth. The suppression of power due to the presence of baryons in
halos is expected to be between $7$ and $15$ per cent, depending on the
feedback model \citep{2013MNRAS.434..148S}. \cite{CFHTLenS-3D} present a
conservative 3D cosmic shear analysis including this model for baryonic
effects.

The Coyote emulation parameters are physical densities $\omega_{\rm m}
= \Omega_{\rm m} h^{-2}$ and $\omega_{\rm b} = \Omega_{\rm b} h^{-2}$.
We sample from those parameters, and calculate $\Omega_{\rm m}$ as
deduced parameter for the final PMC sample. The prior range in
$\Omega_{\rm m}$ and $\sigma_8$ given by the emulator is relatively
narrow.  This makes fitting a power-law $\sigma_8 \Omega_{\rm
  m}^\alpha$ to the degeneracy direction difficult, resulting in an
under-estimated value of $\alpha$ of around $0.4$.  Instead, we fix
$\alpha = 0.713$, which we obtained by sampling the full prior range
of $\Omega_{\rm m}$ and $\sigma_8$ using the revised \texttt{halofit}
model \citep{2012ApJ...761..152T}. The results are given in Table
\ref{tab:Sigma_8_comp}. Note that the smaller prior range results in
smaller error bars compared to the full parameter range.  The
difference between the models is only about 20\% of the statistical
uncertainty.

\begin{table}
  \caption{Marginalised 68.3\% constraints for the amplitude parameter
    $\Sigma_8 = \sigma_8 (\Omega_{\rm m} / 0.27)^{0.713}$ 
    using the CFHTLenS aperture-mass dispersion. We compare two
    models of the non-linear power spectrum. The power-law index $\alpha = 0.713$ is fixed.
    The prior range in both cases is the domain of the Coyote
    emulator, with $\Omega_{\rm m} \in [0.18; 0.48]$ and $\sigma_8 \in [0.6; 0.9]$.}

\renewcommand{\arraystretch}{1.5}
\begin{tabular}{@{}llc} \hline
  Model      & Reference                    & $\sigma_8 (\Omega_{\rm m} / 0.27)^{0.713}$ \\ \hline
  Coyote     & \cite{2014ApJ...780..111H}  & $0.792^{+0.038}_{-0.045}$ \\
  revised \texttt{halofit} & \cite{2012ApJ...761..152T} & $0.785^{+0.038}_{-0.045}$ \\ \hline
\end{tabular}
\renewcommand{\arraystretch}{1.0}
  \label{tab:Sigma_8_comp}
\end{table}

 \begin{figure}
   \resizebox{0.9\hsize}{!}{  
     \includegraphics[bb = 80 50 380 550]%
     {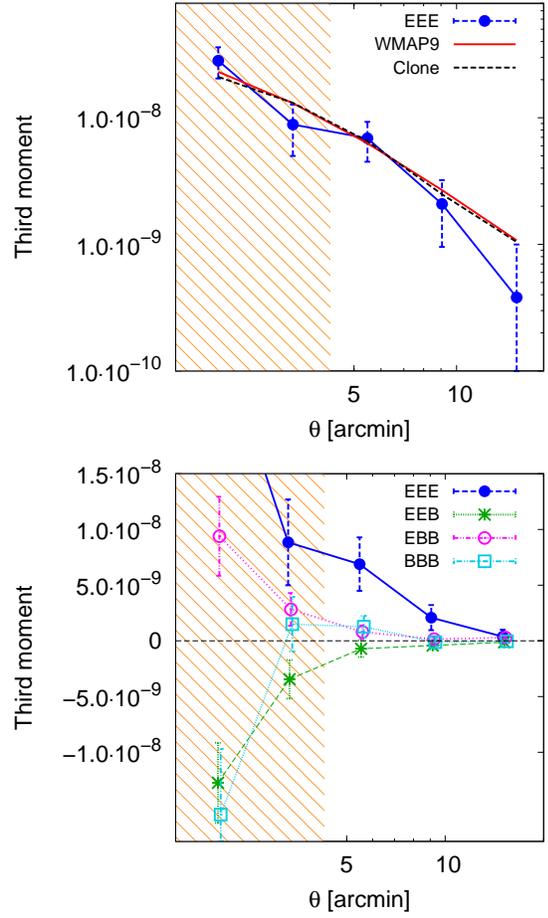}
   }

   \caption{The third-oder aperture-mass E- and B-mode components as
     function of smoothing scale $\theta$, measured from CFHTLenS
     data. \emph{Upper panel:}  The EEE component is shown as
     blue filled circles.  The prediction from WMAP9 is shown as red
   solid line, the third moment measured from the Clone is the black
   dash-dotted curve. \emph{Lower panel:} The B-mode components
   ($EEB$: green crosses; $EBB$: magenta circles; $BBB$: cyan
   squares), measured from the full mosaic data.  The shaded scales
   are not used for cosmological constraints.}

   \label{fig:map3G}

 \end{figure}

 \begin{figure*}  
 \resizebox{0.7\hsize}{!}{  \includegraphics[bb = 140 50 535 250]{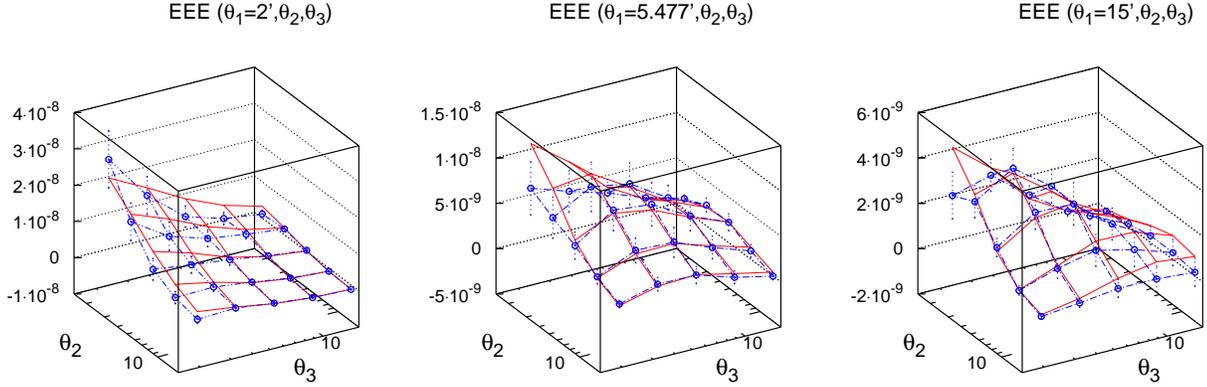}}

 \caption{The generalised third-order aperture-mass E-mode
   $EEE(\theta_1, \theta_2, \theta_3)$ measured from the CFHTLenS
   mosaic catalog (blue surface, with open circle) is compared to the
   prediction from WMAP9 (red surface). In each panel, one angular
   scale $\theta_1$ is fixed, from left to right: $\theta_1 =
   2\arcmin, 5.477\arcmin, 15\arcmin$.  }

   \label{fig:map3GEEE}

 \end{figure*}

 \begin{figure*}  
 \resizebox{0.7\hsize}{!}{  \includegraphics[bb = 140 50 535 250]{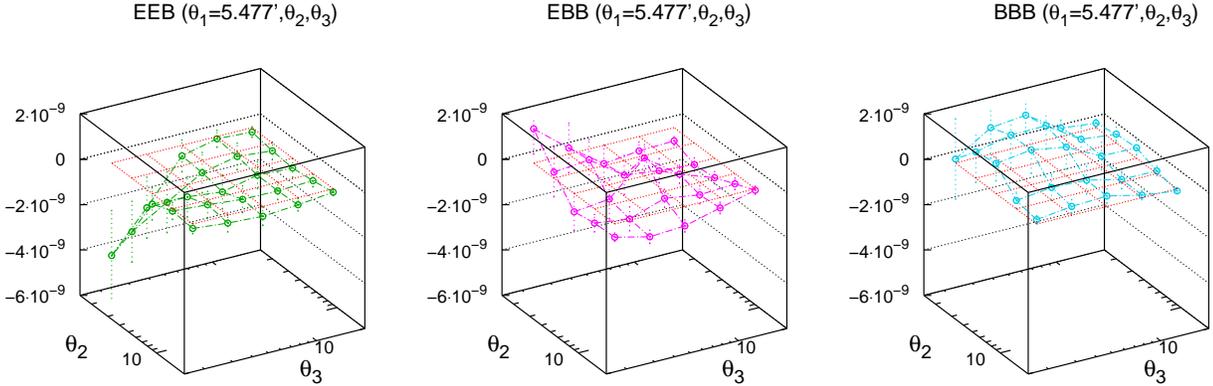}}
 \caption{The generalised third-order aperture-mass B-mode compoments $EEB(\theta_1,
       \theta_2, \theta_3)$ (\emph{left panel}),
   $EBB(\theta_1, \theta_2,
       \theta_3)$ (\emph{middle}),
   $BBB(\theta_1, \theta_2,
       \theta_3)$ (\emph{right}), measured from the
   CFHTLenS mosaic catalog are shown for one fixed radius $\theta_1 =
   5.477\arcmin$. The red grid is the zero surface.}
   \label{fig:map3GBone}
 \end{figure*}

\subsection{Third-order measures}
\label{sec:results_third}

The upper panel of Fig.~\ref{fig:map3G} shows the third-order
aperture-mass moment measured from CFHTLenS data.  This is the
diagonal part for three equal filter scales $\theta$ in the range 2 to
15 arcmin. The plot shows the E-mode ($EEE$; eq.~\ref{Map3}). The
three B-mode components (\ref{Mx3} - \ref{MapMxMx}) are shown on a
linear scale in the lower panel.  All error bars are calculated from
the 184 independent Clone fields of view, rescaled to the observed
survey area, and contain Poisson noise and cosmic variance.

There is good agreement of the E-mode signal with the theoretical
model using the WMAP9 best-fit parameters, and the measurement from
the Clone simulations.  We note a non-zero B-mode detection. The
smallest scale of 2 arcmin shows two non-zero B-mode data points. This
scale may suffer from numerical integration imprecisions due to the
small number of available triangles. Further, intrinsic alignment may
create a B-mode signal on small scales \cite[][hereafter
  SHvWS08]{2008MNRAS.388..991S}. More thorough tests of systematics of
the third-order aperture-mass moment is performed in the companion
paper \cite{CFHTLenS-3pt}. On larger scales the $BBB$ component is
non-zero.  This component is not parity-invariant and is only produced
when the observed shear field shows a parity violation.  We discuss
possible origins of this contribution in Sect.~\ref{sec:discussion}.

A further consistency check is the comparison of the third moment from
the mosaic catalogue to the one measured on single MegaCam pointings
individually. To obtain the error bars of the latter, we subdivide the
Clone fields into $3\times 3$ parts, to account for the smaller
observed field size. This results in larger error bars, in particular
on large angular scales, where substantially fewer triangles are
available. As can be seen in Fig.~\ref{fig:map3G}, the two methods of
obtaining the third moment are consistent. As expected, the greatest
differences occur on large scales, where the relative number of common
triangles is smallest.

Figures \ref{fig:map3GEEE} and \ref{fig:map3GBone} show examples of
the generalised third-order aperture-mass components for a few
combinations of angular smoothing radii ($\theta_1, \theta_2,
\theta_3$).  Except on the smallest scale, the agreement of the
CFHTLenS third-moment E-mode with WMAP9 predictions are very good.
The B-mode is non zero for a few data points, similar to the diagonal
case as discussed above.

\subsection{E- and B-mode measurement significance}
\label{sec:null_tests}

We perform $\chi^2$ null tests of the various E-/B-mode
components. The $\chi^2$ function is given as the Gaussian
distribution (\ref{log_likeli}) where the model $y$ is zero
everywhere.  Thus, the full Clone covariance is taken into account for
the significance test, accounting for the correlation between angular
scales. Contrary to the E-mode, the B-modes covariance only contains
shape noise and no cosmic variance, since there is no cosmological
B-mode signal in the clone. The error bars on the B-mode are therefore
much smaller than for $EEE$. Since there is no intrinsic alignment in
the Clone simulations, the cosmic variance from this contribution is
therefore not included in our covariance, which might over-estimate
the $\chi^2$ significance.

\begin{figure}
  \begin{center}
  \resizebox{0.8\hsize}{!}{
    \includegraphics{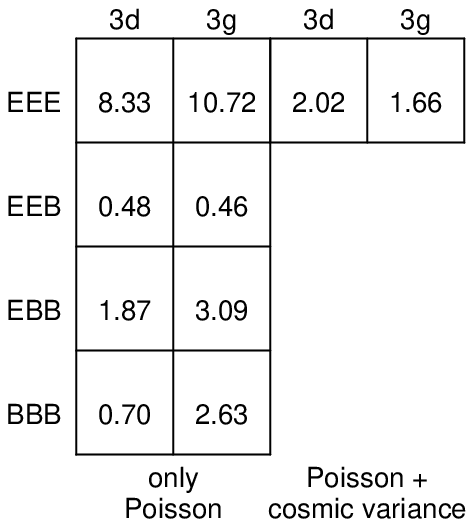}
  }
  \end{center}
  \caption{Each box shows the value $\sigma$ of a significance test,
    where the number in the box denotes the significance in
    $\sigma$. The first and third (second and fourth) columns
    correspond to the diagonal (generalised) third moments, and are
    labeled `3d' (`3g'). The first two columns use Poisson error only,
    the last two columns also include cosmic variance (which is not
    present in the Clone for B-mode components).  The first row is the
    cosmological detection significance, for which higher numbers are
    better.  All subsequent rows are null tests, for which smaller
    numbers are better.}
  \label{fig:null_plot}
\end{figure}

As for the cosmological analysis, we use scales between $5.5$ and $15$
arcmin.  We also check the consistency of the E-mode signal with
theory, in which case the assumed model $y$ is the WMAP9
prediction. Given the degrees of freedom 3 for the diagonal and 10 for
the general third moment, the resulting $\chi^2$ is translated into a
significance level. The results are shown in Fig.~\ref{fig:null_plot}.

The significance of the E-mode is about $2\sigma$ when we include the cosmic
variance. Using Poisson noise only, 
we obtain a much higher significance of more than $8\sigma$.
This covariance would be the correct one to use in case
of absence of $EEE$, since in this case there would be no cosmic variance.
Thus, we can reject the hypothesis of a null third-order lensing signal
with $9\sigma$.

The $EEE$ signal is in very good agreement with the WMAP9 best-fit
model. All diagonal B-mode components are less significant than the
E-mode, and their amplitude is below the E-mode. However, both the
generalised $EEB$ and generalised $BBB$ components are non-null at
about $3\sigma$.

At this time we do not know the origin of those B-modes. Further
speculations are presented in Sect.~\ref{sec:discussion-EB}. Note that
for the joint CFHTLenS+CMB constraints, presented in
Sect.~\ref{sec:joint_constraints}, we only use the diagonal third
moment, for which the B-mode significance is lower.

\subsection{Cosmological constraints}
\label{sec:cosmo_constraints}

We test two predictions about third-order weak lensing statistics: (1)
The generalised third-order aperture-mass moment contains more
information about cosmology than the `diagonal' term \citep{SKL05,
  KS05}. (2) Combined with second-order, parameter degeneracies are
partially lifted, leading to significantly improved joint constraints
\citep{2004MNRAS.348..897T, KS05}. We have already explored these two
predictions using the CFHTLenS Clone simulations
(Sect.~\ref{sec:scales_clone}).

\begin{figure}
  \begin{center}

  Diagonal and generalised third moment
  \resizebox{0.9\hsize}{!}{
    \includegraphics[bb=0 0 360 310]{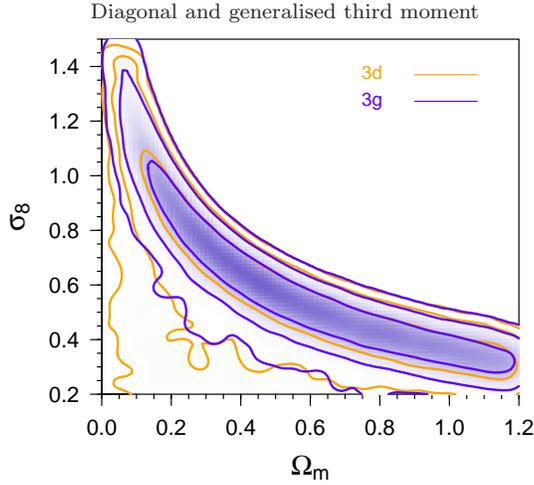}
  }
  \end{center}

  \caption{Marginalised posterior density contours (68.3\%, 95.5\%,
    99.7\%) for the CFHTLenS third-order aperture-mass. The diagonal
    third moment (`3d'; orange lines) is compared to the generalised third
    moment (`3g'; blue contours). The model is flat $\Lambda$CDM.  }

  \label{fig:LCDM_cfhtls_3diag+3gen}

\end{figure}

In Fig.~\ref{fig:LCDM_cfhtls_3diag+3gen}, we show the marginalised constraints
for $\Omegam$ and $\sigma_8$, the parameters that are best constraints from
weak cosmological lensing. Symbols used in the following figures are explained
in Table \ref{tab:symbols}. The generalised third-order aperture-mass covers
indeed a smaller part of parameter space compared to the diagonal one.  Adding
the non-equal smoothing scale measurements of the generalised third moment
rules out those models with a very low $\sigma_8$ and $\Omegam$.   The
amplitude parameter $\Sigma_8$ is larger than zero at more than $3\sigma$ for
both the diagonal and general third moment. This is at much higher significance
than the non-zero detection of $EEE$ (previous section). This result is
stronger, since it involves parameter fitting within the framework of an
assuming theoretical model. In particular, the shape of the signal plays a role
and adds information that is not used in a simple $\chi^2$ null test. This result
is consistent with what we see in the Clone simulations.

\begin{table}
\caption[]{Second- and third-order measures and the corresponding
  symbols used in plots.%
}

\begin{center}

\begin{tabular}{@{}lp{23em}@{}} \hline
Symbol & Description \\ \hline
2      & $\langle M_{\rm ap}^2 \rangle(\theta)$, aperture-mass
dispersion \\
$\mathit{2}$      & COSEBis, a second-order E-/B-mode measure \citep{COSEBIs} \\
3d$\,\,$%
\footnote{Note that the symbol `3d' indicates the third-moment diagonal, and
is not to be confused with three-dimensional (3D) lensing,
e.g.~\cite{CFHTLenS-3D}.}
     & $\langle M_{\rm ap, diag}^3 \rangle = \langle M_{\rm ap}^3 \rangle(\theta)$, diagonal third-order aperture-mass
moment, evaluated for one filter scale \\
3g     & $\langle M_{\rm ap, gen}^3 \rangle = \langle M_{\rm ap}^3 \rangle(\theta_1, \theta_2, \theta_3)$, generalised third-order aperture-
mass moment, correlating three filter scales \\
SLC    & (diagonal) third-order aperture-mass moment from source-lens
clustering (Sect.~\ref{sec:SLC}) \\
IA     & (diagonal) third-order aperture-mass moment from intrinsic
alignments (Sect.~\ref{sec:IA}) \\
\hline
\end{tabular}

\end{center}
\label{tab:symbols}
\end{table}

\begin{figure}  
  \centerline{flat $\Lambda$CDM}

 \resizebox{\hsize}{!}{
   \includegraphics[bb = 0 20 350 310]{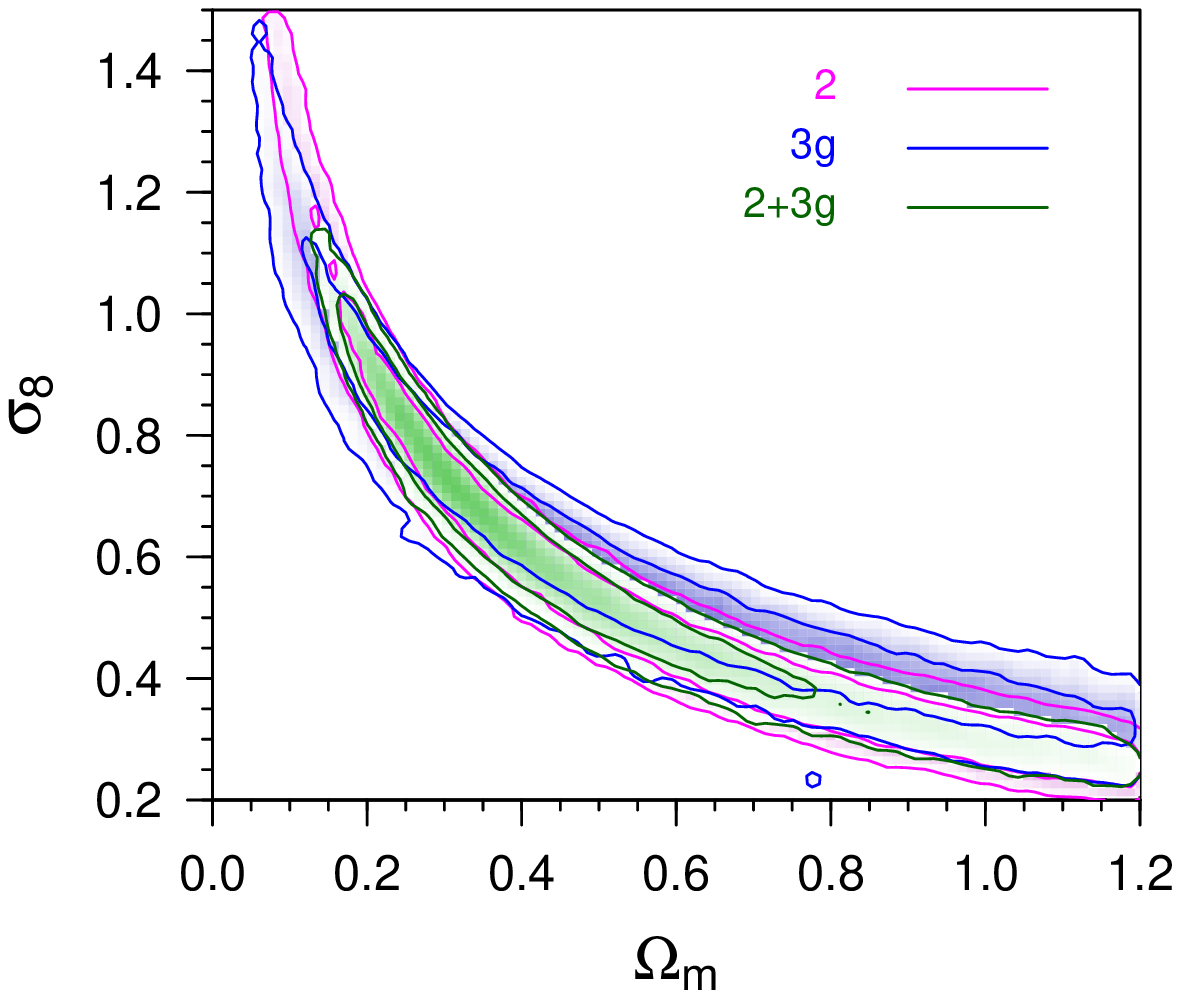}
 }

  \bigskip
  \centerline{flat $\Lambda$CDM}

 \resizebox{\hsize}{!}{
   \includegraphics[bb = 0 20 350 310]{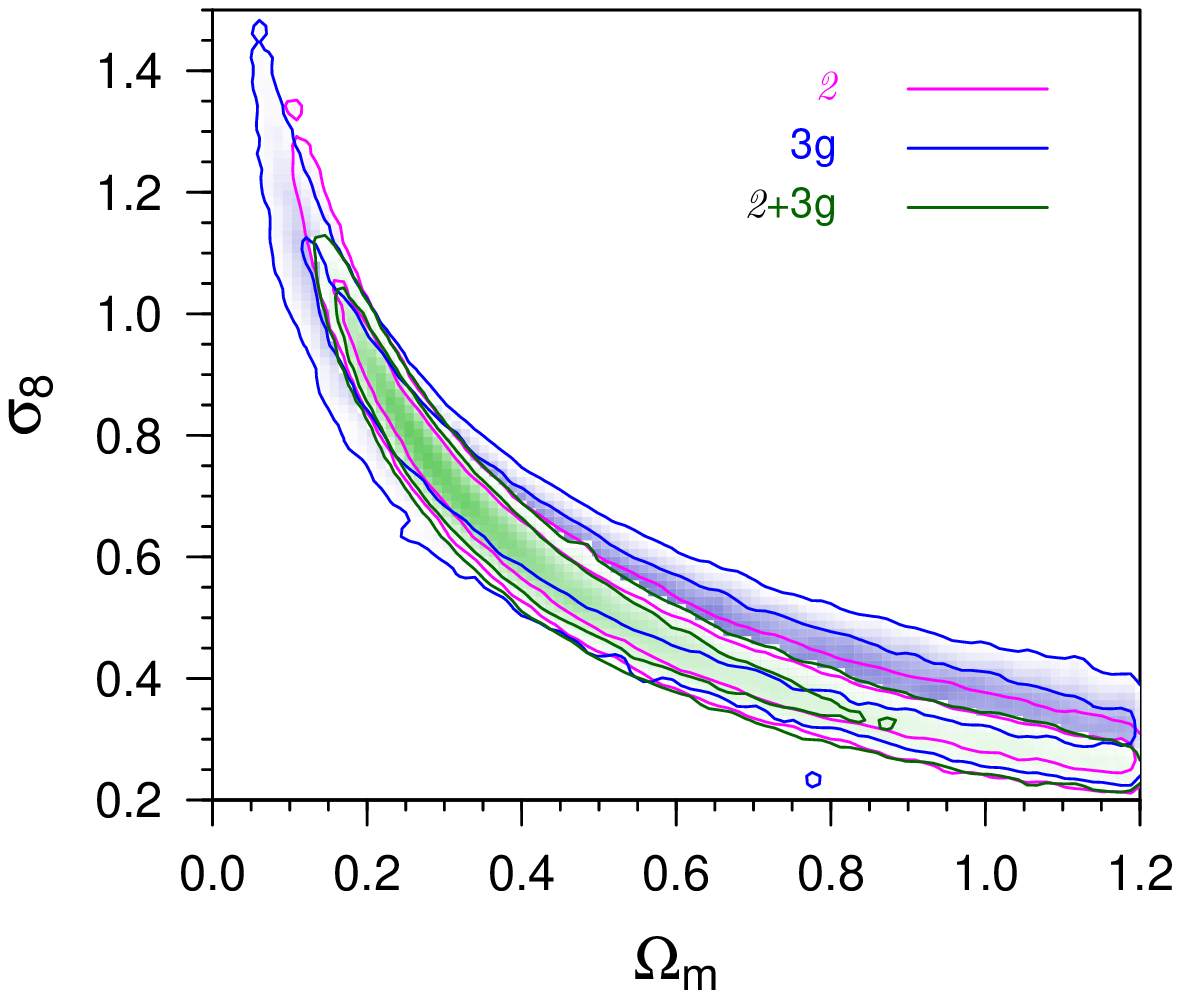}
 }

 \caption{Marginalised posterior density contours (68.3\%, 95.5\%,
   99.7\%) for $\Omega_{\rm m}$ and $\sigma_8$ from
   CFHTLenS. Second-order statistics (magenta contours) are the
   aperture-mass dispersion (\emph{top panel}) and the COSEBIs
   (\emph{bottom}). The blue contours correspond to the generalised
   third-order aperture-mass.  Both second- and third-order measures
   are combined to yield joint constraints (green). The model is flat
   $\Lambda$CDM.}

   \label{fig:LCDM_cfhtls}
 \end{figure}

Second, adding second-order measures reduces the allowed parameter
space, however not by much as shown in Fig.~\ref{fig:LCDM_cfhtls}.
Third-order lensing probes a shallower slope $\alpha$ of the parameter
$\Sigma_8 = \sigma_8 (\Omega_{\rm m}/0.27)^\alpha$, in agreement with
the theoretical prediction from \citet{KS04} and
\cite{2010APh....32..340V}.  Mainly in the region of extreme $\Omegam$
and $\sigma_8$ is where the $\langle M_{\rm ap}^2 \rangle$- and
$\langle M_{\rm ap}^3 \rangle$-constraints differ.  The constraints
orthogonal to the $\Omega_{\rm m}$-$\sigma_8$ degeneracy direction are
reduced by $10$\% ($40$\%) when adding third-order to COSEBIs
(aperture-mass dispersion). Here we see an example where a Fisher
matrix analysis \citep{2004MNRAS.348..897T, KS05} can provide overly
optimistic predictions \citep{WKWG12}. Even though the slope of the
constraints at the fiducial model is different, the curved, non-linear
shape of the parameter degeneracy directions of the two probes largely
negates this difference, leading to a larger overlap between the
allowed regions.  This shows the necessity to explore the full
likelihood function, in our case with Monte-Carlo sampling, to obtain
realistic joint constraints.

\begin{figure}
  \centerline{flat $w$CDM}

\resizebox{\hsize}{!}{

   \includegraphics[bb = 0 20 350 310]{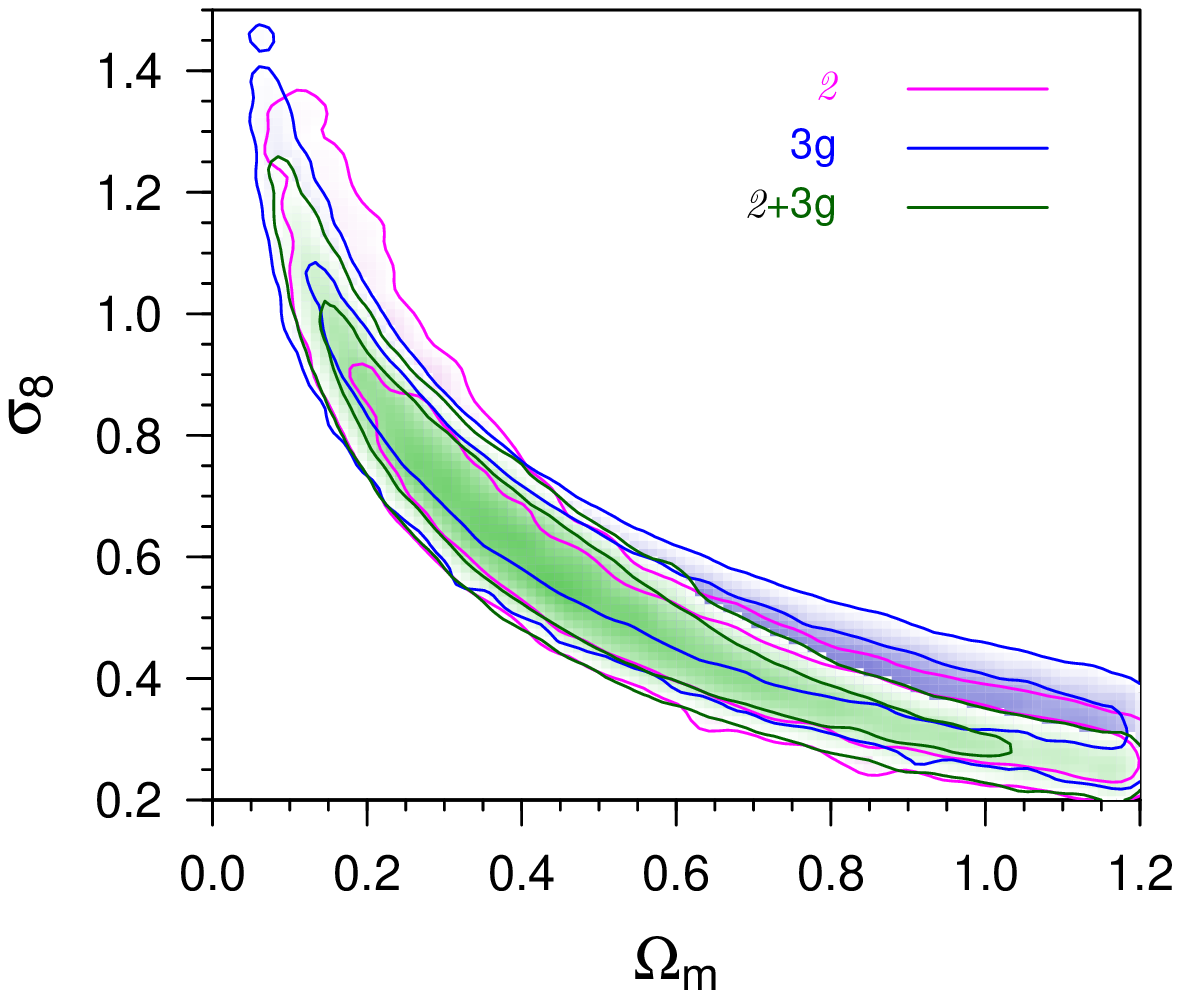}
 }

  \bigskip
  \centerline{curved $\Lambda$CDM}

\resizebox{\hsize}{!}{
   \includegraphics[bb = 0 20 350 310]{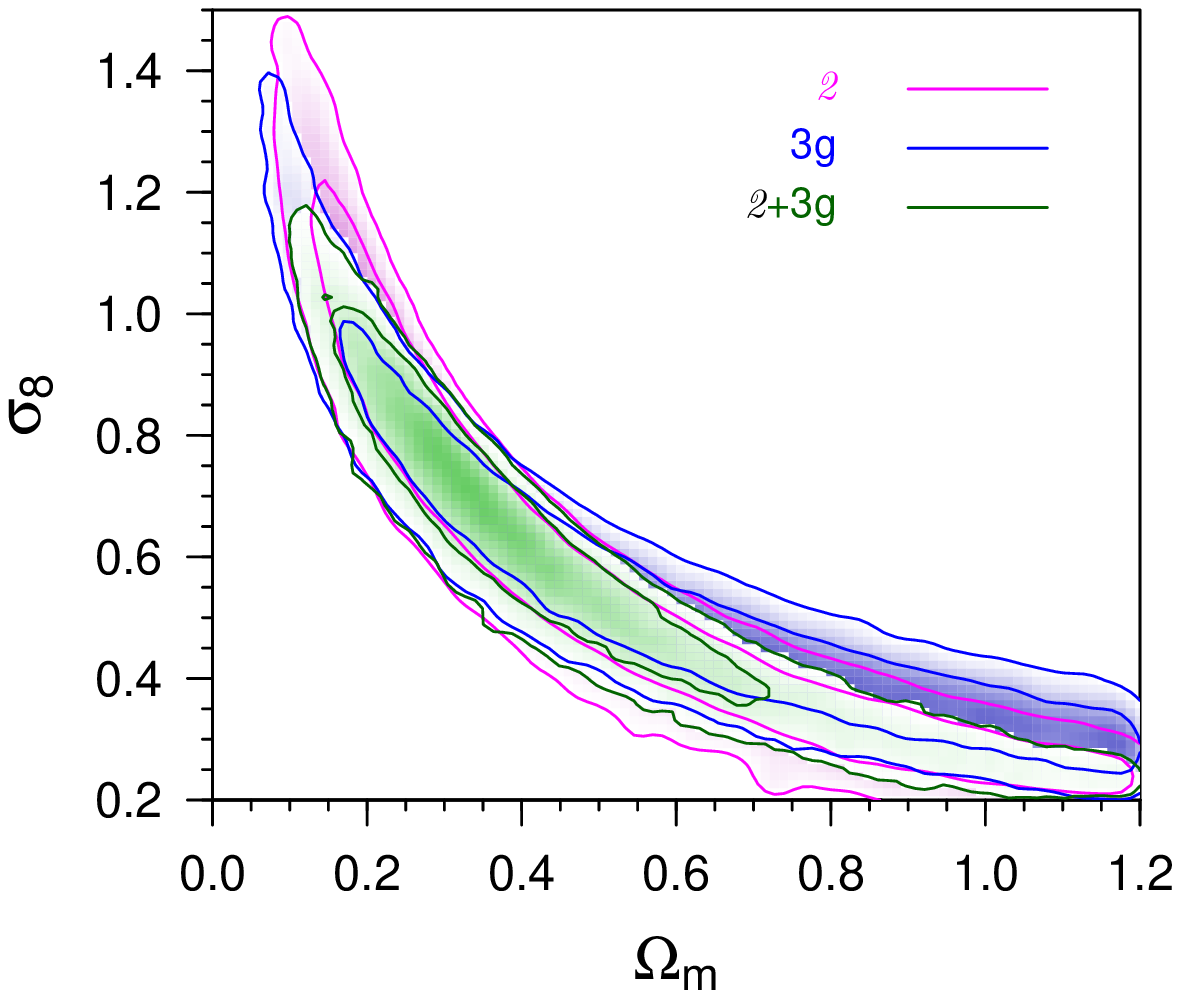}
 }

 \caption{Marginalised posterior density contours (68.3\%, 95.5\%,
   99.7\%) for $\Omega_{\rm m}$ and $\sigma_8$ from
   CFHTLenS. Second-order measures (magenta curves) and third-order
   generalised aperture-mass (blue) are combined to yield joint
   constraints (green). A flat $w$CDM Universe in used the top panel,
   and a curved $\Lambda$CDM model in the bottom panel. }

   \label{fig:Sig8vsOmegam_Lensing}

\end{figure}

We explore extensions from the standard model $\Lambda$CDM model, by
adding (1) curvature, and (2) dark energy in the form of a constant
equation-of-state parameter $w$. For those extensions, the results on
$\Omegam$ and $\sigma_8$ are similar to the standard case, see
Fig.~\ref{fig:Sig8vsOmegam_Lensing}. For further parameters, we
combine CFHTLenS with other probes, see
Sect.~\ref{sec:joint_constraints}.

\section{Astrophysical weak-lensing contaminants}
\label{sec:contaminants}

Third-order cosmic shear statistics suffer from two major contaminants
of astrophysical origin: Intrinsic alignment (IA) and source-lens
clustering (SLC).  Contrary to second-order statistics, IA and SLC
contribute to a much higher level.  The correlations they introduce
can be comparable in amplitude to the cosmological weak-lensing
skewness.

\subsection{Source-lens clustering}
\label{sec:SLC}

Source-lens clustering (SLC), see \cite{1998A&A...338..375B, H02},
denotes the fact that galaxies in a weak-lensing survey act both as
sources and lenses. More precisely, source galaxies are correlated to
structures that cause the lensing effect on other source galaxies. For
a given line of sight, this clustering gives rise to a modulation of
the lensing signal, since the source redshift distribution is changed
with respect to the average in a way that correlates with the lensing
signal. This introduces an additional variance, skewness, etc. of the
convergence field.

To model SLC, we have to use a locally-varying source galaxy density
$p(\vec \theta, w)$ instead of the mean distribution $\bar p(w)$,
which are related to each other as
\begin{equation}
  p(\vec \theta, w) = \bar p(w) \left[ 1 + \delta_{\rm g}(\vec \theta, w) \right].
  \label{plocal}
\end{equation}
We assume a simple linear, deterministic galaxy bias and write
\begin{equation}
  \delta_{\rm g}(\vec \theta, w) = b(w) \delta(\vec \theta, w).
\end{equation}
Inserting the modified galaxy distribution (\ref{plocal}) into
(\ref{kappa}) leads to higher-order density correlations. These
additional terms do not contribute more than a few percent to
second-order statistics, since they are proportional to the
convergence bispectrum. They are however of more importance for the
third-order functions: both cosmic shear, for common models such as
perturbation theory or HEPT, and SLC depend on terms that are
proportional to the power spectrum squared.  Inserting that additional
convergence term into (\ref{map_defx}) yields the SLC contribution to
the aperture-mass,
\begin{align}
  M_{\rm ap}^{\rm SLC}(\theta, \vec \vt) = & \int \dd^2 \vt^\prime \,
  U_\theta(|\vec \vt - \vec \vt^\prime|)
  \nonumber \\
  & \times
  \int_0^{w_{\rm lim}} \dd w  \, \bar 
  p(w) b(w) \delta(\vec \vt, w) \kappa(\vec \vt,
  w) .
  \label{map_slc-2}
\end{align}

Note that we are not estimating the aperture-mass third moment from
local measures of $M_{\rm ap}$, for example by placing apertures over
the survey and then computing the third moment of that
distribution. Instead, we integrate over the 3PCF, which has been
globally computed by averaging over all galaxy triples.  Any local
estimator would need to be normalised by the number of galaxies in
that region, e.g. the aperture disk. That would include the
SLC-corrected $p(w, \vec \theta)$, partly off-setting the SLC
contribution. In a perturbative ansatz, this is represented by two
contributing terms with opposite signs \citep{1998A&A...338..375B}.
Our global estimators of the 3PCF (\ref{gamma0}, \ref{gamma1}) are
instead normalised by the number of galaxy triples over the whole
survey. The SLC correction to that is very small compared to the
expectation value of the unperturbed number of triples. Therefore, we
can safely neglect this contribution \citep{2014A&A...561A..53V}.

We write the total aperture-mass as $M_{\rm ap}^{\rm tot} = M_{\rm ap}
+ M_{\rm ap}^{\rm SLC}$, and expand the third moment up to linear terms in
the SLC contribution (\ref{map_slc-2}). The result is
\begin{align}
    \left\langle \left( M_{\ap}^{\rm tot} \right)^3 \right\rangle (\theta)
    =
    \left\langle M_{\ap}^3 \right\rangle (\theta)
    + 3 \left\langle M_{\ap}^2 M_{\rm ap}^{\rm SLC} \right\rangle (\theta) + \ldots
\label{totskew}
\end{align}
The second term is the first-order SLC contribution. Inserting
(\ref{map_slc-2}), this term can be written as
\begin{align}
  \MoveEqLeft
  3 \left\langle M_{\ap}^2 M_{\rm ap}^{\rm SLC} \right\rangle (\theta)
  = \, 9 \, \Omega_{\rm m} \left(\frac{H_0}{c}\right)^3
  \nonumber \\
  & \times
    \int_0^{w_{\rm lim}} \frac{\dd w}{f_K^2(w)} \, G(w) \bar p(w) \, b(w)
    \nonumber \\
    & \times
  \int_0^w \frac{\dd w^\prime}{a(w^\prime)} \, G(w^\prime) f_K(w - w^\prime)
  \nonumber \\ 
  & \times
  \int_0^\infty \frac{\dd \ell \ell}{2\pi} \hat U(\theta \ell) 
    P_\delta\left(\frac{\ell}{f_K(w)}, w\right)
    \nonumber \\
    & \times
    \int_0^\infty \frac{\dd \ell^\prime \ell^\prime}{2\pi} \hat U(\theta \ell^\prime)
   P_\delta\left(\frac{\ell^\prime}{f_K(w^\prime)}, w^\prime\right)
  Q(\ell \theta, \ell^\prime \theta),
\end{align}
with
\begin{align}
  \MoveEqLeft
  Q(y, y^\prime) = \int_0^{2\pi} \frac{\dd \beta}{2\pi}
    \hat U\left(|\vec y + \vec y^\prime|\right)
    \nonumber \\
      = & \, 2 y^2 {y^\prime}^2 \hat U(y) \hat U(y^\prime)
        \left[ \left( \frac{1}{y^2} + \frac{1}{{y^\prime}^2} \right) {\rm I}_0(y y^\prime)
      - \frac 2 {y y^\prime} {\rm I}_1(y y^\prime) \right],
  \label{Q_SLC}
\end{align}
where ${\rm I}_\nu$ is the modified Bessel function of order
$\nu$. This expression corresponds to the first term of equation (A17)
in \cite{H02}. The latter was obtained for a top-hat filter function,
for which the two Fourier integrals separate. In our case of a
compensated Gaussian filter, the closed expression (\ref{Q_SLC})
describes the mode coupling.

For the galaxy bias we take the redshift-scaling from \cite{1998MNRAS.299...95M},
\begin{equation}
  b(w) = 1 + (b_0 - 1) / D^\gamma_+(w).
  \label{bias_SLC}
\end{equation}
where $D_+$ is the linear growth factor. This implies a bias of unity
at high redshift, which well matches our magnitude-limited sample
\citep{Giannantonio+13}. At $z=0$ the bias is $b_0$.

\subsection{Intrinsic alignment}
\label{sec:IA}

To model the contamination of galaxy intrinsic alignment (IA) to
third-order cosmic shear, we implement the model from SHvWS08.  This
work measured IA in a suite of $N$-body simulations, and modelled IA
for different redshift ranges and a few simple galaxy populations.

For second-order cosmic shear with a very broad source redshift
distribution as is the case in this work, intrinsic alignment plays a minor
role. Its contamination of the cosmological lensing signal is of the order per
cent \citep{CFHTLenS-2pt-tomo}, which we neglect here. See \cite{CFHTLenS-IA}
for a measurement of second-order IA using narrow redshift bins. For
third-order shear however, the contribution from IA is much larger, and we will
model it as follows.

Third-order galaxy shape correlations including shear ($G$) and
intrinsic shape ($I$) can be written schematically as a sum of the
four terms $\langle GGG \rangle + \langle GGI \rangle + \langle GII
\rangle + \langle III \rangle$. The first term, $GGG$, is a pure
lensing correlation. This is the quantity from which we deduce
cosmological parameters. The last term, $III$, is a pure intrinsic
shape correlation. We can safely neglect this term, since only
physically close galaxies give rise to this correlation. For our very
wide redshift bin, the number of such close triples is very small
compared to the overall number of galaxy triples.  The expected
  $III$ amplitude is more than an order of magnitude smaller than GGG,
  an in fact consistent with zero for CFHTLS-type surveys (SHvWS08).

The mixed terms, stemming from the correlation between intrinsic shape
and shear, are produced by galaxy triples at all redshift ranges, and
can be very large compared to $GGG$.  The redshift scaling of these
terms is easy to calculate, since it only depends on the geometry of
the Universe.  For the angular scaling, we follow
SHvWS08 and assume a simple power-law dependence
for the third-order aperture-mass.

Following \citet{2004PhRvD..70f3526H}, the redshift-dependence of the
shear-shape (GI) correlation is straight-forwardly calculated. The
lensing of a source galaxy at redshift $z_{\rm s}$ by structures
correlated to a galaxy at lens redshift $z_{\rm l}$ scales as
$f_K[w(z_{\rm s}) - w(z_{\rm l})] / f_K[w(z_{\rm s})]$. For GGI and
GII, we take into account the redshifts of the galaxy triple, and
integrate over the redshift distributions, neglecting the clustering
of galaxies as a higher-order contribution.  Using a simple
exponential scaling with angular distance \citep[][
SHvWS08]{2005A&A...441...47K} we obtain
\begin{align}
  M_{GII} & = A_{GII} \times \exp\left({\theta / \theta_{\rm GII}}\right)
  \int\limits_0^{z_{\rm lim}} {\rm d} z_{\rm l} \, p^2(z_{\rm l})
  \nonumber \\
  & \times
  \int\limits_0^{z_{\rm l}} {\rm d} z_{\rm s} \, p(z_{\rm s})
  \frac{f_K[w(z_{s}) - w(z_l)]}{f_K[w(z_{s})]};
  \nonumber \\
  M_{GGI} & = A_{GGI} \times \exp \left(\frac{\theta}{\theta_{\rm GGI}}\right)
 \int\limits_0^{z_{\rm lim}} {\rm d} z_{\rm l} \, p(z_{\rm l})
  \nonumber \\
  & \times
  \int\limits_0^{z_{\rm l}} {\rm d} z_{\rm s_1} \, 
  \int\limits_0^{z_{\rm l}} {\rm d} z_{\rm s_2} \, 
  \prod\limits_{i=1}^2 p(z_{\rm s_i}) \frac{f_K[w(z_{s_i}) - w(z_l)]}{f_K[w(z_{s_i})]} .
\end{align}
The IA model parameters are the amplitudes $A_{GII}, A_{GGI}$ and the
characteristic angular scales $\theta_{GII}$ and $\theta_{GGI}$.

We add $M_{GGI}$ and $M_{GII}$ to our theoretical third-order
aperture-mass, and try to jointly sample cosmological and IA
parameters. Due to the relatively low statistical significance of the
CFHTLenS weak-lensing skewness and very limited redshift resolution,
we do not aim to obtain interesting constraints on very general IA
parameters. Rather, our goal is to use a realistic IA model to assess
the influence on our cosmological results.

We therefore use the results from SHvWS08 as priors on our IA
parameters. We use two models of the galaxy population: A realistic
one (mixed early- and late-types) and a pessimistic case (early-types
only).  The redshift combinations tested in SHvWS08 closest to the
CFHTLenS range correspond to the case of lens galaxies at $z_{\rm l} < 1$ and
source galaxies at $z_{\rm s} = 1$. This corresponds roughly to our
mean source redshift of $\bar z = 0.75$, and lens redshifts probed by
a single redshift bin. The best-fit values and error bars for the four
IA parameters are given in Table \ref{tab:IA_priors}. We translate
those into Gaussian priors with width equal to three times the
$1\sigma$ error, while we exclude  unphysical negative scales
$\theta_{GGI}$ and $\theta_{GII}$.

\subsection{Baryonic physics}
\label{sec:baryons}

The presence of baryons in dark-matter halos in the form of stars and
gas changes halo properties {compared to pure dark matter. This} has
an influence on the total power spectrum and bispectrum on small and
medium scales. Prescriptions to quantify and model this, e.g.~with a
halo-model approach, have been obtained by using hydro-dynamical
$N$-body simulations \cite[e.g.][]{2006ApJ...640L.119J,
  2008ApJ...672...19R, 2011MNRAS.415.3649V,2011MNRAS.417.2020S}. The
effect depends on the assumed details of baryonic physics. In the most
realistic case, the amplitude of the third-order aperture moment at
$5.5$ arcmin is suppressed by $10$-$15$ per cent compared to dark
matter only \citep{2013MNRAS.434..148S}. Contrary to IA or SLC, the
relative effect strongly decreases towards larger scales. At $15$
arcmin, the dark-matter only prediction is biased high by less than
$5$ per cent.

We do not include a model of baryonic effects for the power- and
bispectrum in this work. Using a simple calculation, where we model
the decrease of $\langle M_{\rm ap}^3 \rangle$ as a function of
angular scale according to Fig.~1 of \cite{2013MNRAS.434..148S}, we
find that $\Sigma_8$ increases by $0.040$ ($0.022$) for the model with
larger (smaller) baryonic suppression. So our value of $\Sigma_8$,
ignoring baryonic suppression, might be biased high by $3.1$ to $5.5$
per cent.

\subsection{Results}
\label{sec:results_CFHTLenS_sys}

\begin{figure}
\begin{center}

  \centerline{flat $\Lambda$CDM}

\resizebox{\hsize}{!}{
  \includegraphics[bb=0 0 360 310]{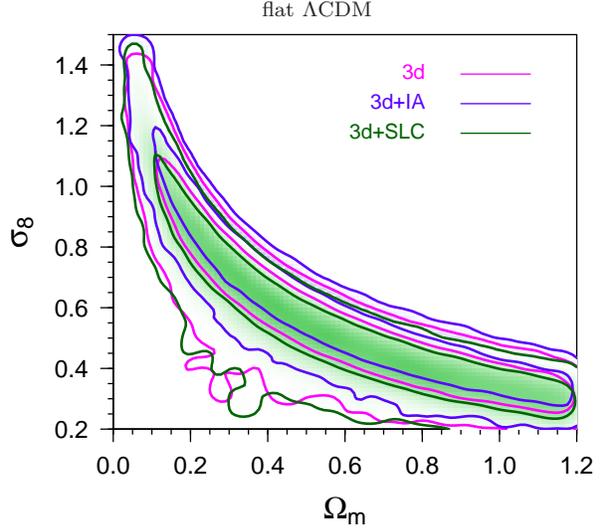}
}

\end{center}

\caption{Marginalised posterior density contours (68.3\%, 95.5\%) for
  $\Omega_{\rm m}$ and $\sigma_8$ from CFHTLenS. We use the
  aperture-mass diagonal third moment which we model with the three
  cases of neglecting astrophysical systematics (`3d'; magenta lines),
  adding intrinsic alignment (`3d+IA'; blue), and source-lens clustering
  (`3d+SLC'; green).}

\label{fig:map3+IA+SLC}

\end{figure}

\begin{table*}
\caption{Intrinsic alignment parameter $1\sigma$ prior ranges from \citet{2008MNRAS.388..991S}.}

\begin{tabular}{@{}lcccc}\hline
Model                    & $A_{GII}/(10^{-7}h\mbox{Mpc}^{-1})$ & $\theta_{GII}/$arcmin & $A_{GGI}/(10^{-7}h\mbox{Mpc}^{-1})$ & $\theta_{GGI}$/arcmin \\
 \hline
mixed (realistic)        & $\phantom{-}0.05 \pm 0.07$  & $1.94 \pm 1.88$ & $-0.15 \pm 0.11$ & $2.36 \pm 1.40$ \\
elliptical (pessimistic) & $-0.04 \pm 0.30$ & $0.37 \pm 2.02$ & $-0.88
\pm 0.09$ & $4.39 \pm 0.38$ \\
\hline
\end{tabular}
\label{tab:IA_priors}
\end{table*}

 \begin{figure}
  \centerline{curved $\Lambda$CDM}

\resizebox{\hsize}{!}{
   \includegraphics[bb = 0 20 350 310]{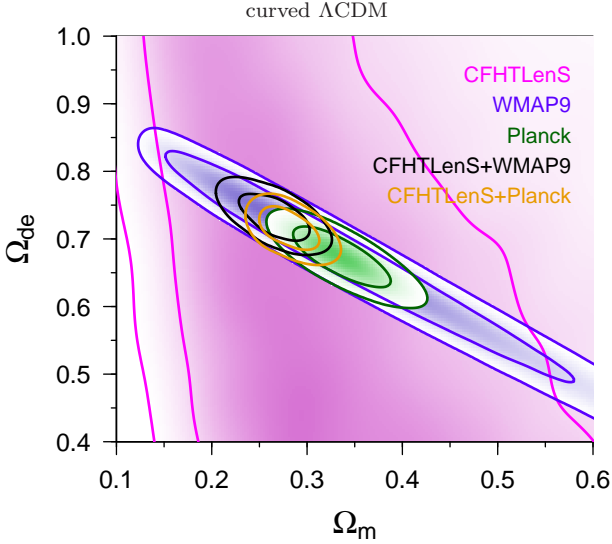}
 }

 \caption{Marginalised posterior density contours (68.3\%, 95.5\%) for
   $\Omega_{\rm m}$ and $\Omega_{\rm de}$ for curved $\Lambda$CDM
   models from CFHTLenS using joint second-order measures (COSEBIs)
   and third-order diagonal aperture-mass (3d) (magenta curves).  The
   constraints from WMAP9 and Planck are shown in blue and green,
   respectively. The joint constraints from CFHTLenS and WMAP9 and
   Planck are shown in black and orange. }

   \label{fig:Sig8vsw0.OmegmvsOmegade}

\end{figure}

Adding IA and SLC changes the amplitude parameter $\Sigma_8$ within
the statistical uncertainty of CFHTLenS. The amplitude change is
comparible in size with the difference between the diagonal and
generalized third moments, see Fig.~\ref{fig:map3+IA+SLC}.

As expected, the total IA contribution ($GGI$ plus $GII$) reduces the
skewness, and the amplitude parameter increases to compensate. There
is only a mild degeneracy between $\sigma_8$ and the IA amplitudes
$A_{GGI}$ and $A_{GII}$. The same is true for the $\Sigma_8$.  The two
IA amplitude parameters are strongly anti-correlated, since they
contribute to the skewness with opposite sign.  The posterior error
bars on the IA amplitude decrease by about 20\% with respect to the
prior.

Using the purely elliptical model from SHvWS08 leads to a very strong
decrease of the third moments. This is the case within the full prior
range, which we take to be three times as large as the $\pm 1\sigma$
errors of SHvWS08. The resulting cosmic shear plus IA aperture-mass
skewness is not compatible with our measurement, and we conclude that
this model is not supported by the data.

SLC leads to a increase of the skewness contrary to IA, and therefore
the jointly fitted $\Sigma_8$ is smaller compared to ignoring this
astrophysical systematic. We explored different bias models and
marginalise over a range of parameters. We found that the results are
not very sensitive within a reasonable model space. The results
presented in this paper correspond to a flat prior with $b \in [0.5;
2]$. The magnitude-limited sample of lensing-selected galaxies between
redshifts of 0.2 and 1.3 is expected to have a mean bias that is not
too far from unity.

When both IA and SLC are included in the joint second- plus
third-order lensing analysis, the resulting amplitude parameter is
marginally increased (Table \ref{tab:sigma_IA}).

\begin{figure*}
{\hspace*{4em} flat $\Lambda$CDM
 \hspace*{22em} curved $\Lambda$CDM}
 \resizebox{\hsize}{!}{ 
 \includegraphics[bb =  0 20 350 310]{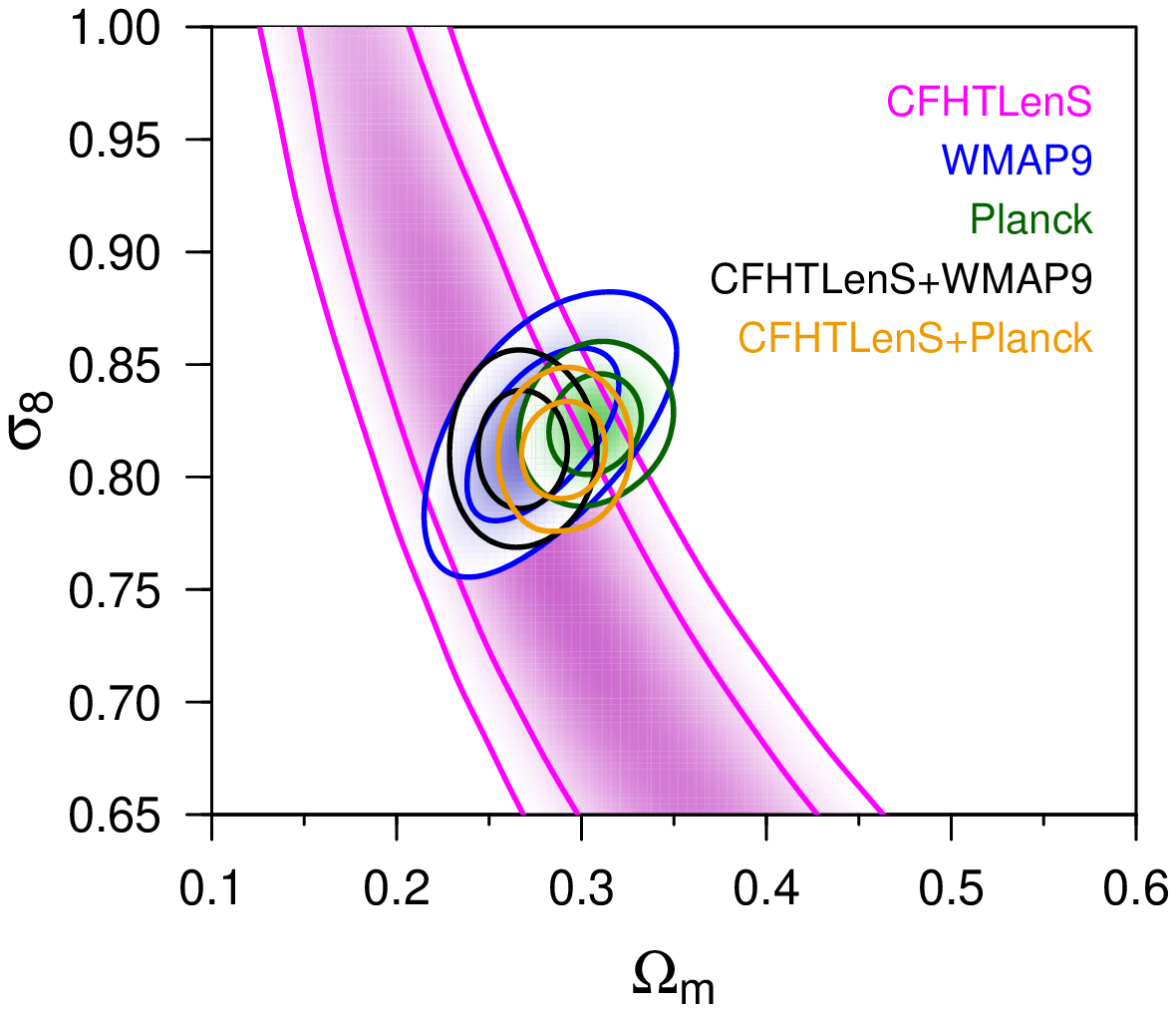}
 \includegraphics[bb =  0 20 350 310]{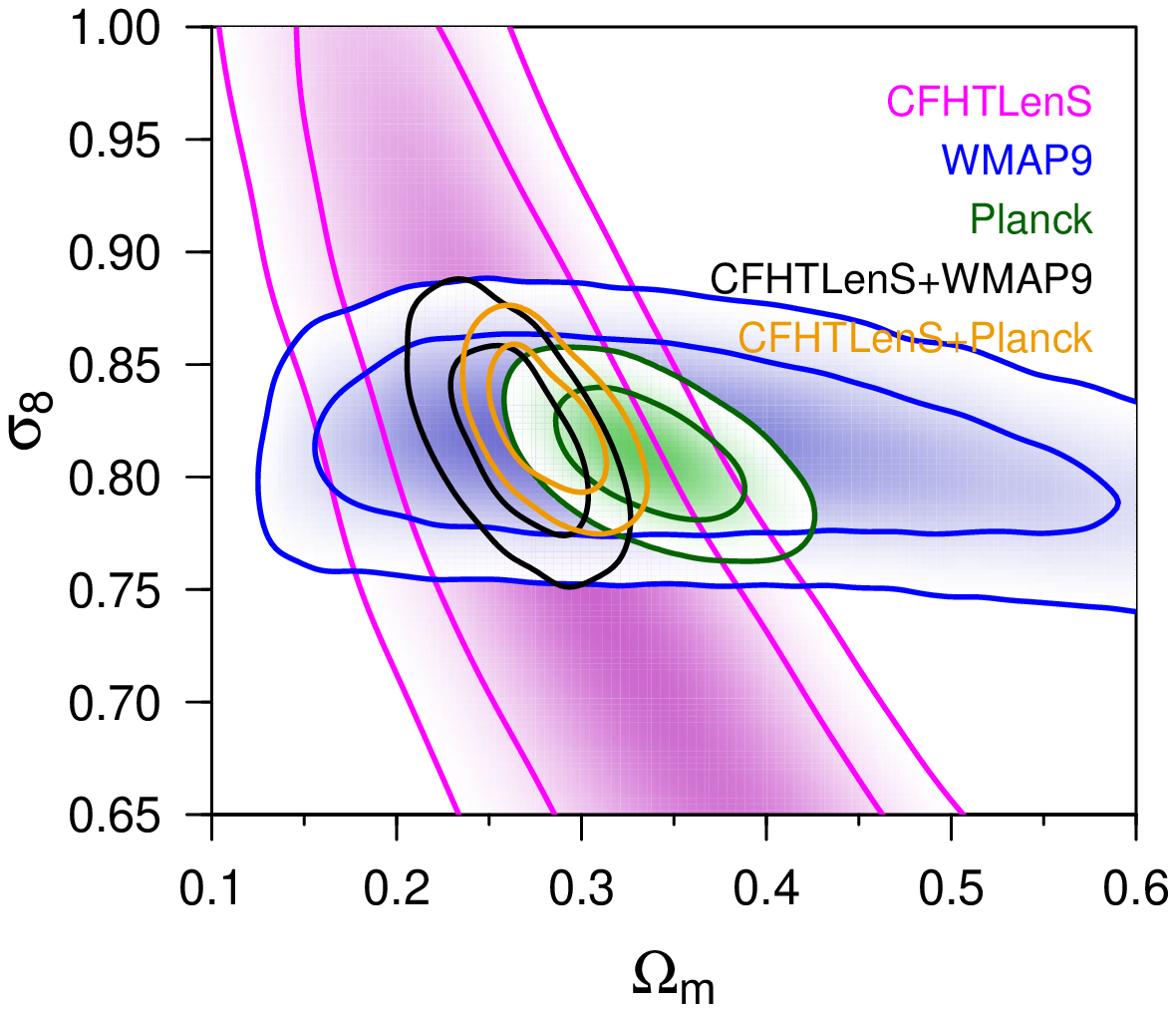}
}
\caption{Marginalised posterior density contours (68.3\%, 95.5\%) from
  CFHTLenS (joint second-order COSEBIs and third-order diagonal
  aperture-mass; magenta lines), WMAP9 (blue), Planck (green),
  CFHTLenS $+$ WMAP9 (black), and CFHTLenS $+$ and Planck
  (orange). The model in the left panel is a flat cosmology, the right
  panel shows the case of free curvature.}
   \label{fig:Sig8vsOmegam_comb} ,
 \end{figure*}

\section{Joint parameter constraints}
\label{sec:joint_constraints}

\begin{table}
  \caption{Constraints from CFHTLenS orthogonal to the
    $\Omegam$-$\sigma_8$ degeneracy direction, using joint
    second-order COSEBIs  and the third-order diagonal aperture-mass
    including SLC and IA. The
    errors are 68\% confidence intervals. The
    three columns correspond to the three different models.}
  \label{tab:sigma_IA}

  \begin{tabular}{|l|l|l|l|l}\hline
    \rule[-3mm]{0em}{8mm}Parameter   & flat\;\;$\Lambda$CDM & flat\;\;$w$CDM	    & curved\;\;$\Lambda$CDM  \\\hline
    $\Sigma_8$ & $ 0.77^{+0.05}_{-0.07} $       & $ 0.77^{+0.09}_{-0.08} $  & $ 0.79^{+0.07}_{-0.12} $   \\
    $\alpha$                         & $ 0.64\pm0.03 $       & $ 0.66\pm0.02 $	    & $ 0.65\pm0.04 $       \\ \hline
  \end{tabular}

\end{table}

We add CMB to CFHTLenS weak-lensing, to lift some of the parameter
constraints.  Throughout this section, the weak-lensing data consists
of the second-order COSEBIs combined with third-order diagonal
aperture mass. For the latter we include intrinsic alignment and
source-lens clustering (Sect.~\ref{sec:contaminants}).  We do not
include a model of the baryonic modification of the power- and
bispectrum. With the conservative choice of angular scales, we limit
the influence of small-scale model unertainties, see also
Sect.~\ref{sec:baryons}.

It is well-known \citep{Contaldi03} and has been recently been
demonstrated again in K13, that the $\Omega_{\rm m}$ - $\sigma_8$
lensing degeneracy in a flat $\Lambda$CDM model can be lifted by the
addition of CMB anisotropies, since the latter shows a near-orthogonal
correlation between those two parameters. In this paper, we combine
second- and third-order CHFLTenS weak lensing with WMAP9 and Planck
measurements. The results are shown in Table~\ref{tab:parameters}.

The Planck constraints are much tighter than the ones from WMAP, and
do therefore dominate the combined confidence region. The Planck error
region is around a slightly higher $\Omega_{\rm m}$ but there is no
significant tension.

For a non-flat cosmology, the difference between WMAP9 and Planck is
much more pronounced, since Planck alone measures the curvature of the
Universe to a high precision due to the addition of CMB lensing
\citep{2006PhRvD..74l3002S}. Moreover, the small size of the joint
constraints for CHFLTenS + Planck comes partially from the fact that
the two confidence regions have a smaller overlap.  We do not consider
this to be a tension: both probes are consistent at 95\% confidence.

The density of the cosmological constant, $\Omega_{\rm de}$ is not
well constrained by 2D lensing.  However, the strong degeneracy for
CMB alone, which is close to the direction of constant curvature, can
be lifted when adding both probes
(Fig.~\ref{fig:Sig8vsw0.OmegmvsOmegade}). This is also true for
Planck, where the degeneracy is smaller due its sensitivity to CMB
lensing, but the joint CHTLenS+Planck constraints are smaller by
$43\%$. Note that the joint CFHTLenS+Planck contour is shifted to
lower values of $\Omega_{\rm m}$ compared to Planck alone, due to the
 differences for $\sigma_8$ between the two probes, see
Fig.~\ref{fig:Sig8vsOmegam_comb}. For a given $\sigma_8$, CFHTlenS
prefers a lower $\Omega{\rm m}$.

\begin{figure}
  \centerline{curved $\Lambda$CDM}

 \resizebox{\hsize}{!}{
   \includegraphics[bb = 0 20 353 310]{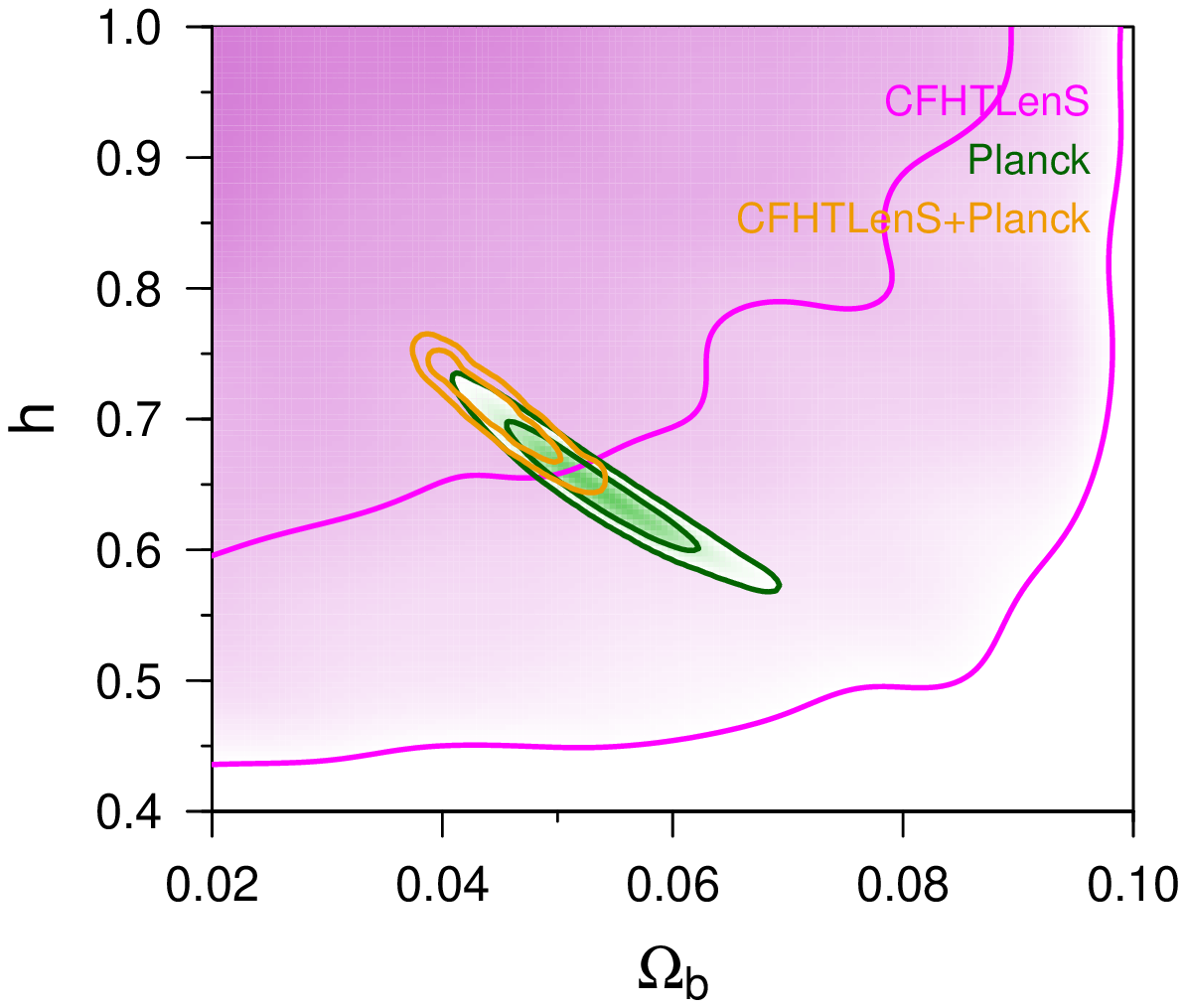}
 }

  \bigskip
  \centerline{curved $\Lambda$CDM}

 \resizebox{\hsize}{!}{
   \includegraphics[bb = 0 20 353 310]{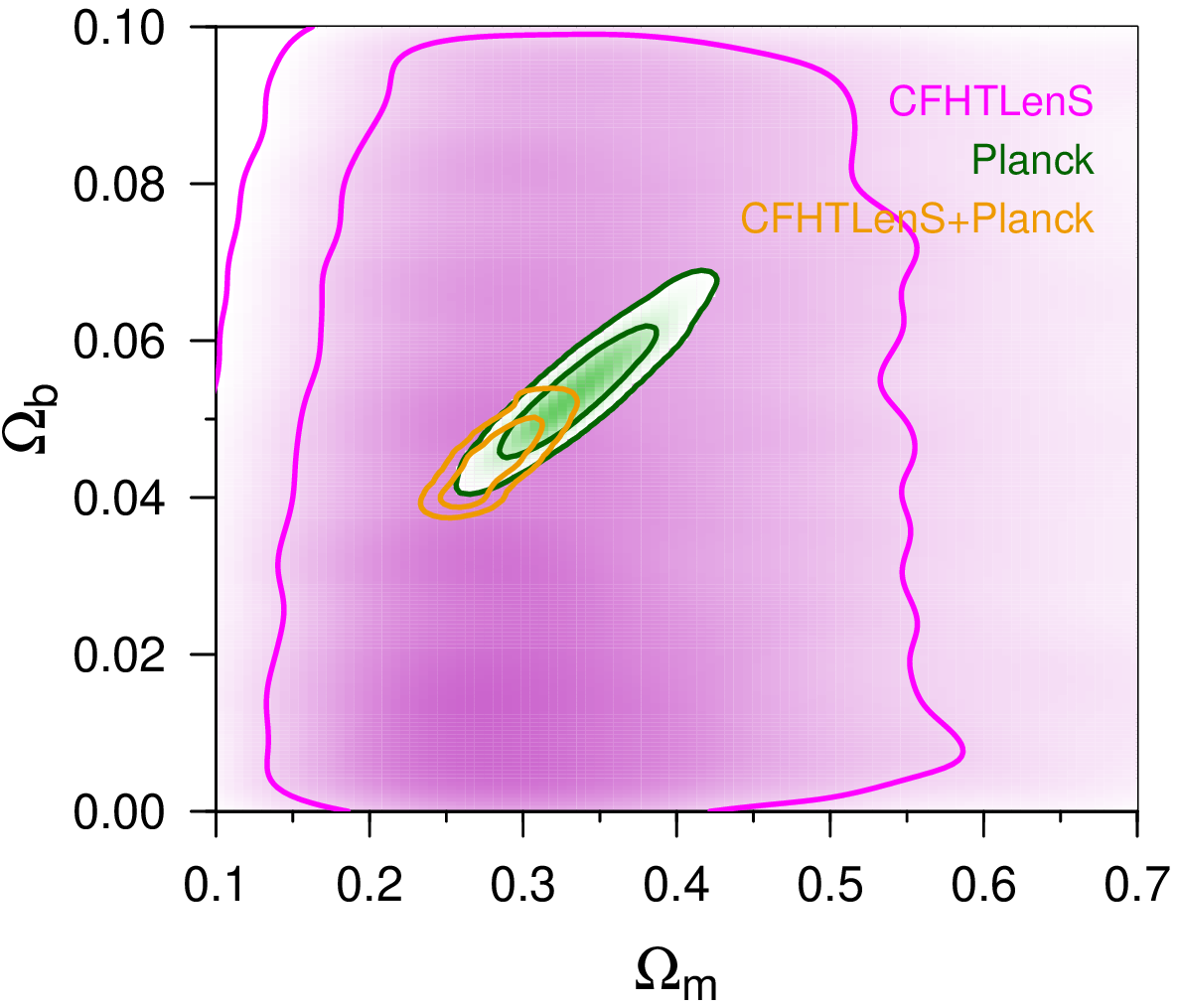}
 }

\caption{ Marginalised posterior density contours (68.3\%, 95.5\%) for
  $\Omega_{\rm b}$ and $h$ (\emph{upper panel}), and $\Omega_{\rm m}$
  and $\Omega_{\rm b}$ (\emph{lower panel}), from CFHTLenS (magenta
  contours), Planck (black lines), and their combination (orange).
  The model is curved $\Lambda$CDM.  }

   \label{fig:Obh_Omh}
 \end{figure}

The CFHTLenS+Planck constraints are shifted towards smaller
$\Omega_{\rm m}$ compared to Planck alone
(Figs.~\ref{fig:Sig8vsw0.OmegmvsOmegade} and
\ref{fig:Sig8vsOmegam_comb}).  It was already seen from second-order
cosmic shear \citep[K13;][]{CFHTLenS-2pt-tomo, CFHTLenS-IA} that
CFHTLenS prefers a lower $\Omega_{\rm m}$.  Further, over the range of
$\Omega_{\rm b}$ allowed by Planck, lensing puts a lower limit on
$h$. Because of the strong CMB degeneracies between $h$ and both
$\Omega_{\rm b}$ and $\Omega_{\rm m}$, the joint Lensing+Planck
constraints rule out larger values of $\Omega_{\rm m}$ and
$\Omega_{\rm b}$ (Fig.~\ref{fig:Obh_Omh}).

For a flat CDM model with a free, constant $w$ parameter, the CMB shows the same
degeneracy as weak lensing in the space of $\Omegam$ and
  $\sigma_8$ as shown in \cite{KB09}. Adding WMAP9 to CFHTLenS does
not reduce the allowed parameter space by much -- to lift the
degeneracy efficiently, one would have to further add BAO and/or
Hubble constant priors.  The combined CFHTLenS+Planck contours are
dominated by Planck. Compared to the flat case, the allowed parameter
space moves towards lower $\Omega_m$ and higher $\sigma_8$.

\begin{table}
  \caption{Cosmological parameter results with 68\% confidence
    intervals. The first line for each
    parameters shows CFHTLenS+WMAP9, the second line is
    CFHTLenS+Planck.
    The three columns correspond to the three
    different models.}
  \label{tab:parameters}

\renewcommand{\arraystretch}{1}
\begin{tabular}{|r|r|r|r|r|}\hline\hline
\rule[-3mm]{0em}{8mm}Param.  &flat\;\;$\Lambda$CDM   &flat\;\;$w$CDM   &curved\;\;$\Lambda$CDM  \\\hline\hline
\  &$ 0.268 ^{+ 0.013 }_{- 0.012 }$    &$ 0.304 ^{+ 0.141 }_{- 0.096 }$    &$ 0.266 ^{+ 0.024 }_{- 0.022 }$   \\
\raisebox{1.5ex}[-1.5ex]{$\Omega_{\textrm{m}}$} &$ 0.290 ^{+ 0.011 }_{- 0.013 }$   &$ 0.187 ^{+ 0.081 }_{- 0.032 }$   &$ 0.282 ^{+ 0.021 }_{- 0.017 }$    \\\hline
\  &$ 0.812 ^{+ 0.014 }_{- 0.013 }$    &$ 0.794 ^{+ 0.148 }_{- 0.120 }$    &$ 0.817 ^{+ 0.028 }_{- 0.024 }$     \\
\raisebox{1.5ex}[-1.5ex]{${\sigma_8}^*$}  &$ 0.812 ^{+ 0.008 }_{- 0.010 }$   &$ 0.975 ^{+ 0.065 }_{- 0.109 }$    &$ 0.823 ^{+ 0.021 }_{- 0.015 }$   \\\hline
\  &\  &$ -0.949 ^{+ 0.355 }_{- 0.466 }$   &\ \\
\raisebox{1.5ex}[-1.5ex]{$w_0$}  &\raisebox{1.5ex}[-1.5ex]{$-1$}   &$ -1.552 ^{+ 0.372 }_{- 0.203 }$  &\raisebox{1.5ex}[-1.5ex]{$-1$}\\\hline
\  &\  &\   &$ 0.733 ^{+ 0.018 }_{- 0.015 }$    \\
\raisebox{1.5ex}[-1.5ex]{$\Omega_{\textrm{de}}$}   &\raisebox{1.5ex}[-1.5ex]{$1-\Omega_{\textrm{m}}$}  &\raisebox{1.5ex}[-1.5ex]{$1-\Omega_{\textrm{m}}$}&$ 0.714 ^{+ 0.012 }_{- 0.016 }$     \\\hline
\  &\  &\  &$ 0.0011 ^{+ 0.0083 }_{- 0.0083 }$    \\
\raisebox{1.5ex}[-1.5ex]{${\Omega_K}$}   &\raisebox{1.5ex}[-1.5ex]{$0$}  &\raisebox{1.5ex}[-1.5ex]{$0$}  &$ 0.0035 ^{+ 0.0074 }_{- 0.0074 }$    \\\hline
\  &$ 0.709 ^{+ 0.013 }_{- 0.013 }$    &$ 0.697 ^{+ 0.168 }_{- 0.116 }$    &$ 0.715 ^{+ 0.042 }_{- 0.032 }$   \\
\raisebox{1.5ex}[-1.5ex]{$h$}  &$ 0.692 ^{+ 0.012 }_{- 0.009 }$    &$ 0.878 ^{+ 0.079 }_{- 0.128 }$  &$ 0.706 ^{+ 0.033 }_{- 0.027 }$     \\\hline
\  &$ 0.0452 ^{+ 0.0013 }_{- 0.0012 }$  &$ 0.0516 ^{+ 0.0240 }_{- 0.0168 }$    &$ 0.0447 ^{+ 0.0048 }_{- 0.0044 }$      \\
\raisebox{1.5ex}[-1.5ex]{$\Omega_{\textrm{b}}$}  &$ 0.0468 ^{+ 0.0009 }_{- 0.0010 }$   &$ 0.0299 ^{+ 0.0127 }_{- 0.0052 }$   &$ 0.0449 ^{+ 0.0037 }_{- 0.0043 }$    \\\hline
\  &$ 0.976 ^{+ 0.012 }_{- 0.012 }$  &$ 0.978 ^{+ 0.014 }_{- 0.013 }$    &$ 0.975 ^{+ 0.011 }_{- 0.012 }$   \\
\raisebox{1.5ex}[-1.5ex]{$n_{\textrm{s}}$}   &$ 0.967 ^{+ 0.009 }_{- 0.005 }$  &$ 0.964 ^{+ 0.006 }_{- 0.006 }$  &$ 0.965 ^{+ 0.009 }_{- 0.006 }$     \\\hline
\end{tabular}

\end{table}

\begin{figure}
\begin{center}
\resizebox{\hsize}{!}{
  \includegraphics[bb=0 0 284 203]{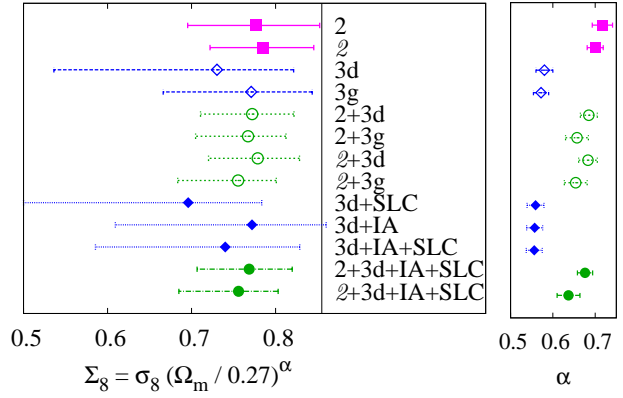}
}
\end{center}

\caption{\emph{Left panel:} 68.3\% constraints for the amplitude parameter $\Sigma_8$.
\emph{Right panel:} Best-fit values of
$\alpha$ with $1\sigma$ errors. The model is a flat $\Lambda$CDM universe.
The numerical values of this graph are given in Table \ref{tab:Sigma}.
See Table \ref{tab:symbols} for a description of
the symbols in the middle column. 
}

\label{fig:Sigma}

\end{figure}

\begin{table}

\caption{Fig.~\ref{fig:Sigma} in table form.  We print for a flat
  $\Lambda$CDM universe the parameter $\Sigma_8 = \sigma_8
  (\Omega_{\rm m} / 0.27)^\alpha$ with mean and $68.3\%$ confidence
  intervals, and the best-fit value with $1\sigma$ error of $\alpha$.
  The probes and their symbols are described in Table
  \ref{tab:symbols}. }

\label{tab:Sigma}

\begin{center}
%
\renewcommand{\arraystretch}{1.4}
\begin{tabular}{|l|l|l|}\hline\hline
\rule[-3mm]{0em}{8mm}$\mbox{Probe}$  &$\Sigma_8$   &$\alpha$  \\\hline\hline
\rule[-1mm]{0em}{5mm}$2$   &$0.78^{+0.08}_{-0.08}$   &$0.72\pm0.02$ \\
${\mathit2}$   &$0.79^{+0.06}_{-0.06}$   &$0.70\pm0.02$ \\
$3\mbox{d}$  &$0.73^{+0.09}_{-0.19}$   &$0.58\pm0.02$ \\
$3\mbox{g}$  &$0.77^{+0.07}_{-0.10}$   &$0.57\pm0.02$ \\
$2+3\mbox{d}$  &$0.77^{+0.05}_{-0.06}$   &$0.69\pm0.02$ \\
$2+3\mbox{g}$  &$0.77^{+0.05}_{-0.06}$   &$0.66\pm0.03$ \\
${\mathit2}+3\mbox{d}$   &$0.78^{+0.05}_{-0.06}$   &$0.68\pm0.02$ \\
${\mathit2}+3\mbox{g}$   &$0.76^{+0.05}_{-0.07}$   &$0.65\pm0.03$ \\
$3\mbox{d}+\mbox{SLC}$   &$0.70^{+0.09}_{-0.20}$   &$0.56\pm0.02$ \\
$3\mbox{d}+\mbox{IA}$  &$0.77^{+0.09}_{-0.16}$   &$0.56\pm0.02$ \\
$3\mbox{d}+\mbox{IA}+\mbox{SLC}$   &$0.74^{+0.09}_{-0.15}$   &$0.56\pm0.02$ \\
$2+3\mbox{d}+\mbox{IA}+\mbox{SLC}$   &$0.77^{+0.05}_{-0.06}$   &$0.68\pm0.02$ \\
${\mathit2}+3\mbox{d}+\mbox{IA}+\mbox{SLC}$  &$0.76^{+0.05}_{-0.07}$   &$0.64\pm0.03$ \\
\hline
\end{tabular}

\end{center}

\end{table}

\section{Discussion}
\label{sec:discussion}

\subsection{Diagonal and generalized third moments}

The generalised third-order aperture-mass contains more information
about cosmology than the diagonal one \citep{KS05}. This is because
the former probes a wide range of triangles of different shape of the
bispectrum in Fourier space. The latter is restricted to mainly
equilateral triangles. This can be inferred from eq.~(\ref{map3-bi}):
if all filter scales $\theta_i$ are equal, the filter functions $\hat
U$ all peak at the same scale $\ell_i = \sqrt(2) / \theta_i$, and the
bispectrum contribution comes mainly from equilateral triangles.

We confirm this prediction with CFHTLenS, where the error bar on the
amplitude parameter $\Sigma_8 = \sigma_8 (\Omega_{\rm m} /
0.27)^\alpha$ is reduced to about half the size (Fig.~\ref{fig:Sigma},
Table \ref{tab:Sigma}). Most importantly, the generalized third moment
excludes a good part of the low-amplitude region in the $\Omega_{\rm
  m}$-$\sigma_8$ parameter space
(Fig.~\ref{fig:LCDM_cfhtls_3diag+3gen}).  Interestingly, despite this
information increase, the detection significance of the generalised
third moment is lower than for the diagonal case
(Fig.~\ref{fig:null_plot}). That shows that the $\chi^2$
  null test is sensitive to different properties of the measurement than
  the Bayesian parameter fit. In particular, there is
  extra-information in the shape of the signal, and not only in the
  amplitude. Only the latter plays a role for the $\chi^2$ null test.

This increase in information for the generalised third moment is also
seen in the Clone simulation, albeit less pronounced. Despite our
attempt to create simulations as realistic as possible, there are
differences which might influence the third-moment measurement in the
Clone in a different way compared to the data. This could be the
noise, but also the underlying cosmological skewness which might be
affected by the finite field and discretisation in the Clone.

The mean value of $\Sigma_8$ is larger for the generalised third
moment compared to the diagonal case. The difference is not
significant due to the relative large error, but we see the same trend
with the Clone simulation.  This may be because the bispectrum model
is not well calibrated on non-equilateral triangles. Another reason
could be that the numerical integration over the 3PCF performs more
poorly for non-equal aperture radii, and gives biased results. This
possibility has to be explored in the future with larger simulations.

\subsection{E- and B-modes}
\label{sec:discussion-EB}

We obtained a $2\sigma$ measurement of the cosmic shear third-order
correlations from CFHTLenS data, but also three cases of B-mode
combinations that are inconsistent with zero at $>1.5\sigma$. One of
these detections, the generalised $BBB$, signifies a violation of
parity invariance. It is not very likely that an astrophysical source
is at the origin of this mode, since it would require a model with a
preferred orientation.

We tested the algorithm to calculate the third-order aperture on the
Clone simulations, and did not find a significant B-mode. It is
however possible that numerical issues are at the root of the B-mode,
such as the discreteness of the triangle binning of the 3PCF, or of
the numerical integration of the 3PCF to obtain the aperture
quantities, or incomplete available integration range.  The fact that
the significance of two of the B-mode components ($EBB$ and $BBB$) is
larger for the generalised third moment is compatible with this view,
since the integration kernels are different for the case with three
different filter scales. More work is however required to scrutinise
this hypothesis.

Undetected PSF correction residuals  could be responsible for
this B-mode.   Third-order PSF-shear correlations were tested and
  the 120 fields used in this work were statistically found to be free
  of systematics \citep{CFHTLenS-3pt}, based on error estimates using
  CFHTLenS Clone simulations, which includes Poisson noise and cosmic
  variance. As in \cite{CFHTLenS-sys}, these tests were done in a
  cosmology-blind way, and therefore did not include E- and B-mode
  decompositions. A full analysis of E- and B-modes from PSF residuals
  is left for future work.

 Further, the lack of intrinsic alignment in the Clone simulations
  might lead to an under-estimation of the covariance matrix, if IA
  produces B-modes, which is very plausible:
  \cite{2013MNRAS.435..194C} obtain E- and B-mode IA power spectrum
  from a generic angular momentum model. SHvWS08 calculate the B-mode
  $III$ term using simple halo-galaxy ellipticity correlations in
  N-body simulations, and find a similar amplitude compared to the
  E-mode. IA generates cosmic variance for the B-mode components,
  which our covariance estimate does not include. This might lead to
  an over-estimation of the B-mode significance. To test this
  hypthesis requires a realistic IA model that produces B-modes.

\subsection{Astrophysical systematics}

We included simple models of two major astrophysical contaminations to
the third-order shear observables.  Both intrinsic alignment and
source-lens clustering depend on the relation of galaxies and dark
matter, and are therefore notoriously difficult to model.

Both IA and SLC modulate the lensing third-order statistics, but the
change on $\Sigma_8$ is smaller than the 68\% statistical error in
both cases.  The error bars on this parameter do not increase when
adding those systematic contributions.

IA consists of two relevant components ($GGI$ and $GII$) with opposite
sign.  Even though the amplitude of each component can be large, their
sum cancels out to a large part. The total contribution to the third
moment is negative. Our adopted galaxy model consists of a mixture of
early- and late-type galaxies. A pure elliptical sample would produce
a very large IA skewness, which is ruled out by our cosmic shear data.

The SLC contribution is proportional to the galaxy bias. We found that
the resulting constraints are not very sensitive to the bias
model. However, for future precision measurements the bias of the
lensing galaxy sample has to be known very well. This is true in
particular for tomographic weak lensing, where galaxies at different
redshifts create an SLC signal that depends on the bias as a function
of redshift.

 We do not include the effect of baryons on the power- and
  bispectrum. With the conservative choice of angular scales, we limit
  the influence of small-scale model unertainties. Following the
  results from $N$-body simulations from \cite{2013MNRAS.434..148S},
  we estimate that the lensing amplitude parameter $\Sigma_8$ from
  third-order alone might be biased high between $3.1$ and $5.5$ per
  cent. 

\subsection{Binning of the three-point correlation function}

We obtain the third-order aperture-mass moment from the shape
catalogue by integrating over the measured and binned three-point
shear correlation function (3PCF). The computation time, even using a
fast tree code, limits us in the use of very small bins. We explore a
range of bin sizes using the Clone simulations, and find convergence
of the results for a reasonable computation time. However, we do see
larger differences in the resulting amplitude of $\langle M_{\rm ap}^3
\rangle$ from the data as a function of bin size
(App.~\ref{sec:binning}).  The differences are well within the
statistical uncertainty of the data, and we leave a more detailed
exploration of the 3PCF calculation for future work.

\subsection{Limitations of cosmology from third-order shear statistic}

Currently, there are limitation for the use of weak-lensing skewness
measures for precision cosmology.  These include:

\begin{itemize}

  \item Model predictions of the bispectrum. Current fitting formulae
    of the bispectrum are accurate to only 10\% on small scales
    \citep{2012JCAP...02..047G}. Moreover, these models have been
    calibrated to only a few basic cosmological models such as
    $\Lambda$CDM. More accurate models are needed, spanning a larger
    parameter space including dark-energy cosmologies.  We tested
    several models in App.~\ref{sec:scales_clone}.

     Further, the effect of baryons has to be studied and
      quantified, for which it is necessary to run large
      hydro-dynamical $N$-body simulations. Predictions based on the
      halo model have already been made for third-order statistics
      \citep{2013MNRAS.434..148S}.  On small scales our lack of
      knowledge of baryonic physics become the dominant uncertainty
      for weak lensing cosmological results.

  \item Astrophysical systematics are not well understood. In
    particular intrinsic alignment models are too simplistic for
    future large-area surveys. Alternatively, IA can be largely
    removed via nulling, at the price of significant loss of
    constraining power \citep{2010A&A...523A..60S}.  SLC is easier
      to model, since it only involves the bias of the lensing galaxy
      sample, which can be obtained with independent observations, for
      example using galaxy clustering.

  \item The Gaussian approximation of the likelihood function will not
    be sufficient for future surveys. Numerical simulations can be
    used to explore and estimate the true likelihood function for a
    limited parameter set \citep{2009A&A...504..689H,
      2009A&A...505..969P}.  In App.~\ref{sec:distribution} we
    calculate the distribution of the third-order aperture-mass
    measurements.

\end{itemize}

In this work we tested and used simple approaches to mitigate those
limitations in the view of current state-of-the-art weak lensing
data. Different models of the bispectrum were compared, intrinsic
alignment was modelled and jointly constrained with cosmology, and the
Gaussianity of the data was measured, and, in the companion paper
\citet{CFHTLenS-3pt}, a non-Gaussian likelihood model was employed.

\section*{Acknowledgments}

The authors thank M.~Jarvis for kindly providing his code to compute
the three-point correlation functions. We acknowledge A.~Grado for
providing the VST cluster at Naples Observatory for computation time,
and P.~Schneider and P.~Valageas for fruitful discussions. We thank the
anonymous referee for useful comments.

This work is based on observations obtained with MegaPrime/MegaCam, a
joint project of CFHT and CEADAPNIA, at the Canada-France-Hawaii
Telescope (CFHT) which is operated by the National Research Council
(NRC) of Canada, the Institut National des Sciences de l'Univers of
the Centre National de la Recherche Scientifique (CNRS) of France, and
the University of Hawaii. This research used the facilities of the
Canadian Astronomy Data Centre operated by the National Research
Council of Canada with the support of the Canadian Space Agency.  We
thank the CFHT staff for successfully conducting the CFHTLS
observations and in particular Jean-Charles Cuillandre and Eugene
Magnier for the continuous improvement of the instrument calibration
and the Elixir detrended data that we used. We also thank TERAPIX for
the individual exposures quality assessment and validation during the
CFHTLS data acquisition period, and Emmanuel Bertin for developing
some of the software used in this study. CFHTLenS data processing was
made possible thanks to significant computing support from the NSERC
Research Tools and Instruments grant program, and to HPC specialist
Ovidiu Toader.

 LF acknowledges support from NSFC grants 11103012 \&11333001,
 Innovation Program 12ZZ134 of SMEC, STCSM grant 11290706600, and
 Shanghai Research grant 13JC1404400.  MJH acknowledges support from
 the Natural Sciences and Engineering Research Council of Canada
 (NSERC). MK is supported in parts by the DFG cluster of excellence
 `Origin and Structure of the Universe'.  CH acknowledges support from
 the European Research Council under the EC FP7 grant number 240185.
 TDK acknowledges support from a Royal Society University Research
 Fellowship.  LVW acknowledges support from the Natural Sciences and
 Engineering Research Council of Canada (NSERC) and the Canadian
 Institute for Advanced Research (CIfAR, Cosmology and Gravity
 program).  TE is supported by the Deutsche Forschungsgemeinschaft
 through project ER 327/3-1 and the Transregional Collaborative
 Research Centre TR 33 - "The Dark Universe".  H. Hildebrandt is
 supported by the Marie Curie IOF 252760, by a CITA National
 Fellowship, and the DFG grant Hi 1495/2-1.  H. Hoekstra acknowledges
 support from Marie Curie IRG grant 230924, the Netherlands
 Organisation for Scientific Research (NWO) grant number 639.042.814
 and from the European Research Council under the EC FP7 grant number
 279396.  YM acknowledges support from CNRS/INSU (Institut National
 des Sciences de l'Univers) and the Programme National Galaxies et
 Cosmologie (PNCG).  BR acknowledges support from the European
 Research Council in the form of a Starting Grant with number 240672.
 MJH acknowledges support from the Natural Sciences and Engineering
 Research Council of Canada (NSERC). TS acknowledges support from NSF
 through grant AST-0444059-001, SAO through grant GO0-11147A, and
 NWO. MV acknowledges support from the Netherlands Organization for
 Scientific Research (NWO) and from the Beecroft Institute for
 Particle Astrophysics and Cosmology.  The $N$-body simulations used
 in this analysis were performed on the TCS supercomputer at the
 SciNet HPC Consortium. SciNet is funded by: the Canada Foundation for
 Innovation under the auspices of Compute Canada; the Government of
 Ontario; Ontario Research Fund - Research Excellence; and the
 University of Toronto.

{\small Author Contribution: All authors contributed to the
  development and writing of this paper. The authorship list consists
  of three alphabetic groups: The first group indicates the two lead
  authors of this paper (LF, MK). The second group includes key
  contributers to the science analysis and interpretation in this
  paper, the founding core team and those whose long-term significant
  effort produced the final CFHTLenS data product. The third group
  covers members of the CFHTLenS team who made a significant
  contribution to the project and/or the paper.  The CFHTLenS
  collaboration was co-led by CH and LVW, and the CFHTLenS Cosmology
  Working Group was led by TK.}

\label{lastpage}

\bibliographystyle{mn2e} \bibliography{astro}

\begin{thebibliography}{81}
\expandafter\ifx\csname natexlab\endcsname\relax\def\natexlab#1{#1}\fi

\bibitem[{Anderson(2003)}]{andersen03}
Anderson T.~W., 2003, An introduction to multivariate statistical analysis, 3rd
  edn. Wiley-Interscience

\bibitem[{{Bartelmann}(2010)}]{2010CQGra..27w3001B}
{Bartelmann} M., 2010, Classical and Quantum Gravity, 27, 233001

\bibitem[{Bartelmann \& Schneider(2001)}]{BS01}
Bartelmann M., Schneider P., 2001, Phys.\ Rep., 340, 297

\bibitem[{{Benjamin} {et~al}\mbox{.}(2013){Benjamin}, {Van Waerbeke},
  {Heymans}, {Kilbinger}, {Erben}, {Hildebrandt}, {Hoekstra}, {Kitching},
  {Mellier}, {Miller}, {Rowe}, {Schrabback}, {Simpson}, {Coupon}, {Fu},
  {Harnois-D{\'e}raps}, {Hudson}, {Kuijken}, {Semboloni}, {Vafaei}, \&
  {Velander}}]{CFHTLenS-2pt-tomo}
{Benjamin} J. {et~al.}, 2013, \mnras, 431, 1547

\bibitem[{{Bernardeau}(1998)}]{1998A&A...338..375B}
{Bernardeau} F., 1998, \aap, 338, 375

\bibitem[{{Bernardeau} {et~al}\mbox{.}(1997){Bernardeau}, {Van~Waerbeke}, \&
  {Mellier}}]{1997A&A...322....1B}
{Bernardeau} F., {Van~Waerbeke} L., {Mellier} Y., 1997, \aap, 322, 1

\bibitem[{{Bernardeau} {et~al}\mbox{.}(2003){Bernardeau}, {Van~Waerbeke}, \&
  {Mellier}}]{2003A&A...397..405B}
{Bernardeau} F., {Van~Waerbeke} L., {Mellier} Y., 2003, \aap, 397, 405

\bibitem[{{Capranico} {et~al}\mbox{.}(2013){Capranico}, {Merkel}, \&
  {Sch{\"a}fer}}]{2013MNRAS.435..194C}
{Capranico} F., {Merkel} P.~M., {Sch{\"a}fer} B.~M., 2013, \mnras, 435, 194

\bibitem[{{Contaldi} {et~al}\mbox{.}(2003){Contaldi}, {Hoekstra}, \&
  {Lewis}}]{Contaldi03}
{Contaldi} C.~R., {Hoekstra} H., {Lewis} A., 2003, Physical Review Letters, 90,
  221303/1

\bibitem[{{Crittenden} {et~al}\mbox{.}(2002){Crittenden}, {Natarajan}, {Pen},
  \& {Theuns}}]{2002ApJ...568...20C}
{Crittenden} R.~G., {Natarajan} P., {Pen} U.-L., {Theuns} T., 2002, \apj, 568,
  20

\bibitem[{{Eifler} {et~al}\mbox{.}(2009){Eifler}, {Schneider}, \&
  {Hartlap}}]{2009A&A...502..721E}
{Eifler} T., {Schneider} P., {Hartlap} J., 2009, \aap, 502, 721

\bibitem[{{Erben} {et~al}\mbox{.}(2013){Erben}, {Hildebrandt}, {Miller}, {van
  Waerbeke}, {Heymans}, {Hoekstra}, {Kitching}, {Mellier}, {Benjamin}, {Blake},
  {Bonnett}, {Cordes}, {Coupon}, {Fu}, {Gavazzi}, {Gillis}, {Grocutt}, {Gwyn},
  {Holhjem}, {Hudson}, {Kilbinger}, {Kuijken}, {Milkeraitis}, {Rowe},
  {Schrabback}, {Semboloni}, {Simon}, {Smit}, {Toader}, {Vafaei}, {van Uitert},
  \& {Velander}}]{CFHTLenS-data}
{Erben} T. {et~al.}, 2013, \mnras, 433, 2545

\bibitem[{{Fu} {et~al}\mbox{.}(2008){Fu}, {Semboloni}, {Hoekstra}, {Kilbinger},
  {Van Waerbeke}, {Tereno}, {Mellier}, {Heymans}, {Coupon}, {Benabed},
  {Benjamin}, {Bertin}, {Dor{\'e}}, {Hudson}, {Ilbert}, {Maoli}, {Marmo},
  {McCracken}, \& {M{\'e}nard}}]{FSHK08}
{Fu} L. {et~al.}, 2008, \aap, 479, 9

\bibitem[{{Giannantonio} {et~al}\mbox{.}(2014){Giannantonio}, {Ross},
  {Percival}, {Crittenden}, {Bacher}, {Kilbinger}, {Nichol}, \&
  {Weller}}]{Giannantonio+13}
{Giannantonio} T., {Ross} A.~J., {Percival} W.~J., {Crittenden} R., {Bacher}
  D., {Kilbinger} M., {Nichol} R., {Weller} J., 2014, \prd, 89, 023511

\bibitem[{{Gil-Mar{\'{\i}}n} {et~al}\mbox{.}(2012){Gil-Mar{\'{\i}}n}, {Wagner},
  {Fragkoudi}, {Jimenez}, \& {Verde}}]{2012JCAP...02..047G}
{Gil-Mar{\'{\i}}n} H., {Wagner} C., {Fragkoudi} F., {Jimenez} R., {Verde} L.,
  2012, \jcap, 2, 47

\bibitem[{{Hamana} {et~al}\mbox{.}(2002){Hamana}, {Colombi}, {Thion},
  {Devriendt}, {Mellier}, \& {Bernardeau}}]{H02}
{Hamana} T., {Colombi} S.~T., {Thion} A., {Devriendt} J.~E.~G.~T., {Mellier}
  Y., {Bernardeau} F., 2002, \mnras, 330, 365

\bibitem[{{Harnois-D{\'e}raps} {et~al}\mbox{.}(2012){Harnois-D{\'e}raps},
  {Vafaei}, \& {Van Waerbeke}}]{CFHTLenS-Clone}
{Harnois-D{\'e}raps} J., {Vafaei} S., {Van Waerbeke} L., 2012, \mnras, 426,
  1262

\bibitem[{{Hartlap} {et~al}\mbox{.}(2009){Hartlap}, {Schrabback}, {Simon}, \&
  {Schneider}}]{2009A&A...504..689H}
{Hartlap} J., {Schrabback} T., {Simon} P., {Schneider} P., 2009, \aap, 504, 689

\bibitem[{{Hartlap} {et~al}\mbox{.}(2007){Hartlap}, {Simon}, \&
  {Schneider}}]{HSS07}
{Hartlap} J., {Simon} P., {Schneider} P., 2007, \aap, 464, 399

\bibitem[{{Heitmann} {et~al}\mbox{.}(2009){Heitmann}, {Higdon}, {White},
  {Habib}, {Williams}, {Lawrence}, \& {Wagner}}]{CoyoteII}
{Heitmann} K., {Higdon} D., {White} M., {Habib} S., {Williams} B.~J.,
  {Lawrence} E., {Wagner} C., 2009, \apj, 705, 156

\bibitem[{{Heitmann} {et~al}\mbox{.}(2014){Heitmann}, {Lawrence}, {Kwan},
  {Habib}, \& {Higdon}}]{2014ApJ...780..111H}
{Heitmann} K., {Lawrence} E., {Kwan} J., {Habib} S., {Higdon} D., 2014, \apj,
  780, 111

\bibitem[{{Heitmann} {et~al}\mbox{.}(2010){Heitmann}, {White}, {Wagner},
  {Habib}, \& {Higdon}}]{CoyoteI}
{Heitmann} K., {White} M., {Wagner} C., {Habib} S., {Higdon} D., 2010, \apj,
  715, 104

\bibitem[{{Heymans} {et~al}\mbox{.}(2013){Heymans}, {Grocutt}, {Heavens},
  {Kilbinger}, {Kitching}, {Simpson}, {Benjamin}, {Erben}, {Hildebrandt},
  {Hoekstra}, {Mellier}, {Miller}, {Van Waerbeke}, {Brown}, {Coupon}, {Fu},
  {Harnois-D{\'e}raps}, {Hudson}, {Kuijken}, {Rowe}, {Schrabback}, {Semboloni},
  {Vafaei}, \& {Velander}}]{CFHTLenS-IA}
{Heymans} C. {et~al.}, 2013, \mnras, 432, 2433

\bibitem[{{Heymans} {et~al}\mbox{.}(2012){Heymans}, {Van Waerbeke}, {Miller},
  {Erben}, {Hildebrandt}, {Hoekstra}, {Kitching}, {Mellier}, {Simon},
  {Bonnett}, {Coupon}, {Fu}, {Harnois D{\'e}raps}, {Hudson}, {Kilbinger},
  {Kuijken}, {Rowe}, {Schrabback}, {Semboloni}, {van Uitert}, {Vafaei}, \&
  {Velander}}]{CFHTLenS-sys}
{Heymans} C. {et~al.}, 2012, \mnras, 427, 146

\bibitem[{{Hildebrandt} {et~al}\mbox{.}(2012){Hildebrandt}, {Erben}, {Kuijken},
  {Van Waerbeke}, {Heymans}, {Coupon}, {Benjamin}, {Bonnett}, {Fu}, {Hoekstra},
  {Kitching}, {Mellier}, {Miller}, {Velander}, {Hudson}, {Rowe}, {Semboloni},
  \& {Ben{\'{\i}}tez}}]{CFHTLenS-photoz}
{Hildebrandt} H. {et~al.}, 2012, \mnras, 421, 2355

\bibitem[{{Hirata} \& {Seljak}(2004)}]{2004PhRvD..70f3526H}
{Hirata} C.~M., {Seljak} U., 2004, \prd, 70, 063526

\bibitem[{{Hoekstra} \& {Jain}(2008)}]{2008ARNPS..58...99H}
{Hoekstra} H., {Jain} B., 2008, Annual Review of Nuclear and Particle Science,
  58, 99

\bibitem[{Jarvis {et~al}\mbox{.}(2004)Jarvis, Bernstein, \& Jain}]{JBJ04}
Jarvis M., Bernstein G., Jain B., 2004, \mnras, 352, 338 (JBJ04)

\bibitem[{{Jee} {et~al}\mbox{.}(2013){Jee}, {Tyson}, {Schneider}, {Wittman},
  {Schmidt}, \& {Hilbert}}]{2012arXiv1210.2732J}
{Jee} M.~J., {Tyson} J.~A., {Schneider} M.~D., {Wittman} D., {Schmidt} S.,
  {Hilbert} S., 2013, \apj, 765, 74

\bibitem[{{Jing} {et~al}\mbox{.}(2006){Jing}, {Zhang}, {Lin}, {Gao}, \&
  {Springel}}]{2006ApJ...640L.119J}
{Jing} Y.~P., {Zhang} P., {Lin} W.~P., {Gao} L., {Springel} V., 2006, \apjl,
  640, L119

\bibitem[{{Kaiser}(1992)}]{1992ApJ...388..272K}
{Kaiser} N., 1992, \apj, 388, 272

\bibitem[{Kaiser {et~al}\mbox{.}(1994)Kaiser, Squires, Fahlman, \&
  Woods}]{KSFW94}
Kaiser N., Squires G., Fahlman G., Woods D., 1994, in Clusters of galaxies,
  Proceedings of the XIVth Moriond Astrophysics Meeting, M\'eribel, France, p.
  269, {also}arXiv:astro-ph/9407004

\bibitem[{{Kayo} {et~al}\mbox{.}(2013){Kayo}, {Takada}, \&
  {Jain}}]{2013MNRAS.429..344K}
{Kayo} I., {Takada} M., {Jain} B., 2013, \mnras, 429, 344

\bibitem[{{Kilbinger} {et~al}\mbox{.}(2009){Kilbinger}, {Benabed}, {Guy},
  {Astier}, {Tereno}, {Fu}, {Wraith}, {Coupon}, {Mellier}, {Balland},
  {Bouchet}, {Hamana}, {Hardin}, {McCracken}, {Pain}, {Regnault}, {Schultheis},
  \& {Yahagi}}]{KB09}
{Kilbinger} M. {et~al.}, 2009, \aap, 497, 677

\bibitem[{{Kilbinger} {et~al}\mbox{.}(2013){Kilbinger}, {Fu}, {Heymans},
  {Simpson}, {Benjamin}, {Erben}, {Harnois-D{\'e}raps}, {Hoekstra},
  {Hildebrandt}, {Kitching}, {Mellier}, {Miller}, {Van Waerbeke}, {Benabed},
  {Bonnett}, {Coupon}, {Hudson}, {Kuijken}, {Rowe}, {Schrabback}, {Semboloni},
  {Vafaei}, \& {Velander}}]{CFHTLenS-2pt-notomo}
{Kilbinger} M. {et~al.}, 2013, \mnras, 430, 2200 (K13)

\bibitem[{Kilbinger \& Schneider(2004)}]{KS04}
Kilbinger M., Schneider P., 2004, \aap, 413, 465

\bibitem[{Kilbinger \& Schneider(2005)}]{KS05}
Kilbinger M., Schneider P., 2005, \aap, 442, 69

\bibitem[{{Kilbinger} {et~al}\mbox{.}(2006){Kilbinger}, {Schneider}, \&
  {Eifler}}]{KSE06}
{Kilbinger} M., {Schneider} P., {Eifler} T., 2006, \aap, 457, 15

\bibitem[{{King}(2005)}]{2005A&A...441...47K}
{King} L.~J., 2005, \aap, 441, 47

\bibitem[{{Kitching} {et~al}\mbox{.}(2014){Kitching}, {Heavens}, {Alsing},
  {Erben}, {Heymans}, {Hildebrandt}, {Hoekstra}, {Jaffe}, {Kiessling},
  {Mellier}, {Miller}, {van Waerbeke}, {Benjamin}, {Coupon}, {Fu}, {Hudson},
  {Kilbinger}, {Kuijken}, {Rowe}, {Schrabback}, {Semboloni}, \&
  {Velander}}]{CFHTLenS-3D}
{Kitching} T.~D. {et~al.}, 2014, submitted to \mnras, also 1401.6842

\bibitem[{{Lawrence} {et~al}\mbox{.}(2010){Lawrence}, {Heitmann}, {White},
  {Higdon}, {Wagner}, {Habib}, \& {Williams}}]{CoyoteIII}
{Lawrence} E., {Heitmann} K., {White} M., {Higdon} D., {Wagner} C., {Habib} S.,
  {Williams} B., 2010, \apj, 713, 1322

\bibitem[{{Miller} {et~al}\mbox{.}(2013){Miller}, {Heymans}, {Kitching}, {van
  Waerbeke}, {Erben}, {Hildebrandt}, {Hoekstra}, {Mellier}, {Rowe}, {Coupon},
  {Dietrich}, {Fu}, {Harnois-D{\'e}raps}, {Hudson}, {Kilbinger}, {Kuijken},
  {Schrabback}, {Semboloni}, {Vafaei}, \& {Velander}}]{CFHTLenS-shapes}
{Miller} L. {et~al.}, 2013, \mnras, 429, 2858

\bibitem[{{Moscardini} {et~al}\mbox{.}(1998){Moscardini}, {Coles}, {Lucchin},
  \& {Matarrese}}]{1998MNRAS.299...95M}
{Moscardini} L., {Coles} P., {Lucchin} F., {Matarrese} S., 1998, \mnras, 299,
  95

\bibitem[{{Munshi} {et~al}\mbox{.}(2008){Munshi}, {Valageas}, {Van~Waerbeke},
  \& {Heavens}}]{2008PhR...462...67M}
{Munshi} D., {Valageas} P., {Van~Waerbeke} L., {Heavens} A., 2008, \physrep,
  462, 67

\bibitem[{Peacock \& Dodds(1996)}]{pd96}
Peacock J.~A., Dodds S.~J., 1996, \mnras, 280, L19

\bibitem[{Peebles(1980)}]{pee80}
Peebles P.~J.~E., 1980, The Large-Scale Structure of the Universe. Princeton
  University Press

\bibitem[{{Pen} {et~al}\mbox{.}(2003){Pen}, {Zhang}, Van~Waerbeke, {Mellier},
  {Zhang}, \& {Dubinski}}]{2003ApJ...592..664P}
{Pen} U.-L., {Zhang} T., Van~Waerbeke L., {Mellier} Y., {Zhang} P., {Dubinski}
  J., 2003, \apj, 592, 664

\bibitem[{{Pires} {et~al}\mbox{.}(2009){Pires}, {Starck}, {Amara},
  {R{\'e}fr{\'e}gier}, \& {Teyssier}}]{2009A&A...505..969P}
{Pires} S., {Starck} J.-L., {Amara} A., {R{\'e}fr{\'e}gier} A., {Teyssier} R.,
  2009, \aap, 505, 969

\bibitem[{{Planck Collaboration} {et~al}\mbox{.}(2014){Planck Collaboration},
  {Ade}, {Aghanim}, {Armitage-Caplan}, {Arnaud}, {Ashdown}, {Atrio-Barandela},
  {Aumont}, {Baccigalupi}, {Banday}, \& et~al.}]{2013arXiv1303.5076P}
{Planck Collaboration} {et~al.}, 2014, \aap\ in press, also 1303.5076

\bibitem[{{Rudd} {et~al}\mbox{.}(2008){Rudd}, {Zentner}, \&
  {Kravtsov}}]{2008ApJ...672...19R}
{Rudd} D.~H., {Zentner} A.~R., {Kravtsov} A.~V., 2008, \apj, 672, 19

\bibitem[{{Sato} \& {Nishimichi}(2013)}]{2013PhRvD..87l3538S}
{Sato} M., {Nishimichi} T., 2013, \prd, 87, 123538

\bibitem[{{Schneider}(1996)}]{1996MNRAS.283..837S}
{Schneider} P., 1996, \mnras, 283, 837

\bibitem[{{Schneider}(2003)}]{2003A&A...408..829S}
{Schneider} P., 2003, \aap, 408, 829

\bibitem[{{Schneider} {et~al}\mbox{.}(2010){Schneider}, {Eifler}, \&
  {Krause}}]{COSEBIs}
{Schneider} P., {Eifler} T., {Krause} E., 2010, \aap, 520, A116

\bibitem[{Schneider {et~al}\mbox{.}(2005)Schneider, Kilbinger, \&
  Lombardi}]{SKL05}
Schneider P., Kilbinger M., Lombardi M., 2005, \aap, 431, 9 (SKL05)

\bibitem[{Schneider \& Lombardi(2003)}]{tpcf1}
Schneider P., Lombardi M., 2003, \aap, 397, 809

\bibitem[{{Schneider} {et~al}\mbox{.}(1998){Schneider}, Van~Waerbeke, {Jain},
  \& {Kruse}}]{1998MNRAS.296..873S}
{Schneider} P., Van~Waerbeke L., {Jain} B., {Kruse} G., 1998, \mnras, 296, 873

\bibitem[{{Schneider} {et~al}\mbox{.}(2002){Schneider}, Van~Waerbeke, \&
  {Mellier}}]{2002A&A...389..729S}
{Schneider} P., Van~Waerbeke L., {Mellier} Y., 2002, \aap, 389, 729

\bibitem[{{Schrabback} {et~al}\mbox{.}(2010){Schrabback}, {Hartlap},
  {Joachimi}, {Kilbinger}, {Simon}, {Benabed}, {Brada{\v c}}, {Eifler},
  {Erben}, {Fassnacht}, {High}, {Hilbert}, {Hildebrandt}, {Hoekstra},
  {Kuijken}, {Marshall}, {Mellier}, {Morganson}, {Schneider}, {Semboloni}, {Van
  Waerbeke}, \& {Velander}}]{SHJKS09}
{Schrabback} T. {et~al.}, 2010, \aap, 516, A63

\bibitem[{{Scoccimarro} \& {Couchman}(2001)}]{2001MNRAS.325.1312S}
{Scoccimarro} R., {Couchman} H.~M.~P., 2001, \mnras, 325, 1312

\bibitem[{{Semboloni} {et~al}\mbox{.}(2008){Semboloni}, {Heymans}, {van
  Waerbeke}, \& {Schneider}}]{2008MNRAS.388..991S}
{Semboloni} E., {Heymans} C., {van Waerbeke} L., {Schneider} P., 2008, \mnras,
  388, 991 (SHvWS08)

\bibitem[{{Semboloni} {et~al}\mbox{.}(2013){Semboloni}, {Hoekstra}, \&
  {Schaye}}]{2013MNRAS.434..148S}
{Semboloni} E., {Hoekstra} H., {Schaye} J., 2013, \mnras, 434, 148

\bibitem[{{Semboloni} {et~al}\mbox{.}(2011{\natexlab{a}}){Semboloni},
  {Hoekstra}, {Schaye}, {van Daalen}, \& {McCarthy}}]{2011MNRAS.417.2020S}
{Semboloni} E., {Hoekstra} H., {Schaye} J., {van Daalen} M.~P., {McCarthy}
  I.~G., 2011{\natexlab{a}}, \mnras, 417, 2020

\bibitem[{{Semboloni} {et~al}\mbox{.}(2011{\natexlab{b}}){Semboloni},
  {Schrabback}, {van Waerbeke}, {Vafaei}, {Hartlap}, \&
  {Hilbert}}]{2011MNRAS.410..143S}
{Semboloni} E., {Schrabback} T., {van Waerbeke} L., {Vafaei} S., {Hartlap} J.,
  {Hilbert} S., 2011{\natexlab{b}}, \mnras, 410, 143

\bibitem[{Semboloni {et~al}\mbox{.}(2014)Semboloni {et~al.}}]{CFHTLenS-3pt}
Semboloni E., {et~al.}, 2014, in prep.

\bibitem[{{Shi} {et~al}\mbox{.}(2010){Shi}, {Joachimi}, \&
  {Schneider}}]{2010A&A...523A..60S}
{Shi} X., {Joachimi} B., {Schneider} P., 2010, \aap, 523, A60

\bibitem[{{Shi} {et~al}\mbox{.}(2014){Shi}, {Joachimi}, \&
  {Schneider}}]{2014A&A...561A..68S}
{Shi} X., {Joachimi} B., {Schneider} P., 2014, \aap, 561, A68

\bibitem[{{Smith} {et~al}\mbox{.}(2006){Smith}, {Hu}, \&
  {Kaplinghat}}]{2006PhRvD..74l3002S}
{Smith} K.~M., {Hu} W., {Kaplinghat} M., 2006, \prd, 74, 123002

\bibitem[{{Smith} {et~al}\mbox{.}(2003){Smith}, {Peacock}, {Jenkins}, {White},
  {Frenk}, {Pearce}, {Thomas}, {Efstathiou}, \&
  {Couchman}}]{2003MNRAS.341.1311S}
{Smith} R.~E. {et~al.}, 2003, \mnras, 341, 1311

\bibitem[{{Takada} \& {Jain}(2003)}]{2003ApJ...583L..49T}
{Takada} M., {Jain} B., 2003, \apjl, 583, L49

\bibitem[{{Takada} \& {Jain}(2004)}]{2004MNRAS.348..897T}
{Takada} M., {Jain} B., 2004, \mnras, 348, 897

\bibitem[{{Takahashi} {et~al}\mbox{.}(2012){Takahashi}, {Sato}, {Nishimichi},
  {Taruya}, \& {Oguri}}]{2012ApJ...761..152T}
{Takahashi} R., {Sato} M., {Nishimichi} T., {Taruya} A., {Oguri} M., 2012,
  \apj, 761, 152

\bibitem[{{Taylor} {et~al}\mbox{.}(2013){Taylor}, {Joachimi}, \&
  {Kitching}}]{2013MNRAS.tmp.1312T}
{Taylor} A., {Joachimi} B., {Kitching} T., 2013, \mnras

\bibitem[{{Vafaei} {et~al}\mbox{.}(2010){Vafaei}, {Lu}, {van Waerbeke},
  {Semboloni}, {Heymans}, \& {Pen}}]{2010APh....32..340V}
{Vafaei} S., {Lu} T., {van Waerbeke} L., {Semboloni} E., {Heymans} C., {Pen}
  U.-L., 2010, Astroparticle Physics, 32, 340

\bibitem[{{Valageas}(2014)}]{2014A&A...561A..53V}
{Valageas} P., 2014, \aap, 561, A53

\bibitem[{{van Daalen} {et~al}\mbox{.}(2011){van Daalen}, {Schaye}, {Booth}, \&
  {Dalla Vecchia}}]{2011MNRAS.415.3649V}
{van Daalen} M.~P., {Schaye} J., {Booth} C.~M., {Dalla Vecchia} C., 2011,
  \mnras, 415, 3649

\bibitem[{{Van Waerbeke} {et~al}\mbox{.}(2013){Van Waerbeke}, {Benjamin},
  {Erben}, {Heymans}, {Hildebrandt}, {Hoekstra}, {Kitching}, {Mellier},
  {Miller}, {Coupon}, {Harnois-D{\'e}raps}, {Fu}, {Hudson}, {Kilbinger},
  {Kuijken}, {Rowe}, {Schrabback}, {Semboloni}, {Vafaei}, {van Uitert}, \&
  {Velander}}]{CFHTLenS-kappa-maps}
{Van Waerbeke} L. {et~al.}, 2013, \mnras, 433, 3373

\bibitem[{{Van~Waerbeke} {et~al}\mbox{.}(1999){Van~Waerbeke}, {Bernardeau}, \&
  {Mellier}}]{1999A&A...342...15V}
{Van~Waerbeke} L., {Bernardeau} F., {Mellier} Y., 1999, \aap, 342, 15

\bibitem[{{Van Waerbeke} {et~al}\mbox{.}(2001){Van Waerbeke}, {Hamana},
  {Scoccimarro}, {Colombi}, \& {Bernardeau}}]{2001MNRAS.322..918V}
{Van Waerbeke} L., {Hamana} T., {Scoccimarro} R., {Colombi} S., {Bernardeau}
  F., 2001, \mnras, 322, 918

\bibitem[{{Wolz} {et~al}\mbox{.}(2012){Wolz}, {Kilbinger}, {Weller}, \&
  {Giannantonio}}]{WKWG12}
{Wolz} L., {Kilbinger} M., {Weller} J., {Giannantonio} T., 2012, \jcap, 9, 9

\bibitem[{{Zaldarriaga} \& {Scoccimarro}(2003)}]{2003ApJ...584..559Z}
{Zaldarriaga} M., {Scoccimarro} R., 2003, \apj, 584, 559

\end{thebibliography}

\begin{appendix}

\section{Clone simulation tests}
\label{sec:clone}

In this appendix we discuss various tests involving the CFHTLenS
`Clone' $N$-body simulations \citep{CFHTLenS-Clone}. These tests
include the validity of the theoretical model, and the accuracy of the
3PCF algorithm.

\subsection{Theoretical model and angular scales}
\label{sec:scales_clone}

\begin{figure}
   \resizebox{0.9\hsize}{!}{
     \includegraphics[bb = 20 10 280 430]{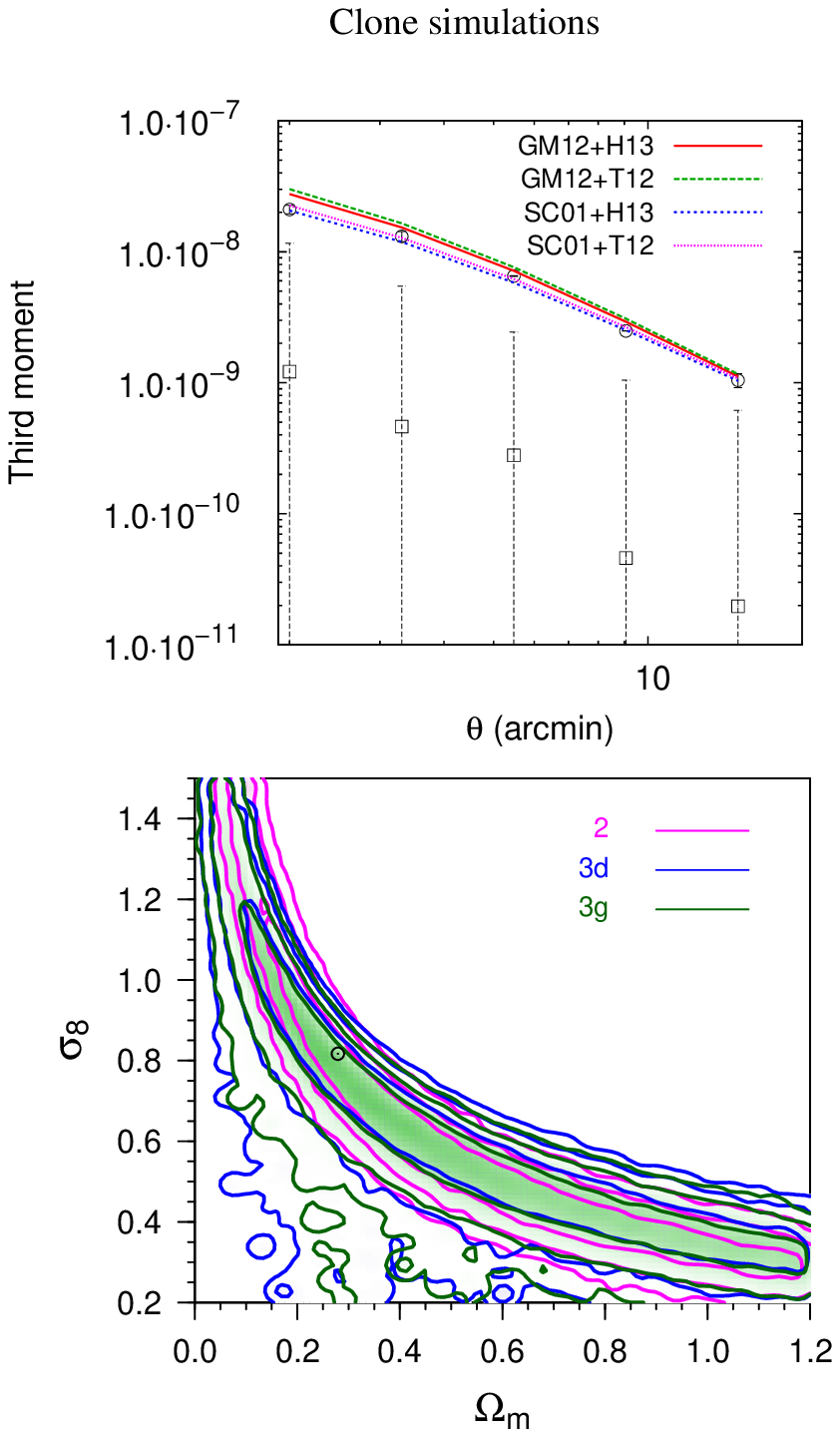}
   }
 
     \caption{\emph{Upper panel:} The third-order aperture-mass predicted
       from Clone
       cosmological parameters using the combinations of bispectrum
       models and non-linear power  models. The models for the power spectrum models are
       T12 \citep{2012ApJ...761..152T} and H13
       \citep{2014ApJ...780..111H}. The bispectrum fitting formulae
       are SC01 \citep{2001MNRAS.325.1312S} and GM12
       \citep{2012JCAP...02..047G}.  The four combinations of models
       are T12+SC01 (solid red lines), T12+GM12 (dashed green),
       H13+SC01 (blue dotted), H13+GM12 (magenta dashed-dotted curve).
         The third moment ($EEE$)  measured from  the Clone are shown as
        open circles, whereas $EBB$ are negative  shown as open squares.
       \emph{Lower panel:} Marginalised posterior
       density contours (68.3\%, 95.5\%, 99.7\%) for $\Omega_{\rm m}$
       and $\sigma_8$ from the Clone mean data vector, using the
       models T12+SC01.  Shown are the disperson (magenta), the diagonal
       third-order aperture-mass (blue curves), and the generalized third order
       (green).  The open circle presents the input cosmology.
          }
   \label{fig:map3model}
 \end{figure}

To check the accuracy of our theoretical model as well as the
numerical algorithms involved to obtain the third-order aperture-mass
moments, we measure $\langle M_{\rm ap}^3 \rangle$ from the Clone
simulation and try to recover the input cosmology using the sampling
method and likelihood function as described in the previous two
sections. We implemented various combinations of prescriptions of the
non-linear power and bi-spectrum as described at the end of
Sect.\ref{sec:cs_theory}. A comparison of those predictions with the
Clone simulations are shown in Fig.~\ref{fig:map3model}.  The
difference between the original halofit \citep{2003MNRAS.341.1311S}
and its revised version is minimal for third order.  The difference
with \citet{2014ApJ...780..111H} is larger but still small compared to
our error bars. Since the validity of the latter is restricted in
parameter space, $k$-, and redshift range, we chose the revised
halofit prescription for the power spectrum. The model from
\citet{2012JCAP...02..047G} overpredicts the Clone simulations, and we
choose \citet{2001MNRAS.325.1312S} for the bispectrum model.

 We try different combinations of minimum and maximum aperture radii
 under the assumption of a flat $\Lambda$CDM model. The parameter
 combination $(\Omegam / 0.279)^\alpha$, which is the direction along
 the $\Omegam - \sigma_8$ degeneracy, is consistent with the input
 value of $\sigma_8 = 0.817$.  The best recovery of the input
 parameters is obtained using the aperture angular range [$5\arcmin;
   15\arcmin]$. We therefore choose this range for the cosmological
 analysis of CFHTLenS, using three filter scales. The number of
 distinct data points for the generalised third moment (combinations
 with repetitions) is 10.

The generalised third moment yields only sightly smaller error bars
compared to the diagonal case, thereby only partially confirming the
predicted strong increase of information from Fisher matrix analyses
\citep{KS05}. In Sect.~\ref{sec:cosmo_constraints} we see that the
CFHTLenS data shows a larger difference between diagonal and
generalised third moment. This is further elaborated in the discussion
(Sect.~\ref{sec:discussion}).

We compare third-order constraints with the aperture-mass dispersion
$\langle M_{\rm ap}^2\rangle (\theta)$, the latter ranging between 2
to 70 arcmin. The direction of degeneracy for $\Omega_{\rm m}$ and
$\sigma_8$ is very similar in both cases, but the slope $\alpha$ of
the elongated ``banana'' $\sigma_8 (\Omega / 0.279)^\alpha$ is
slightly steeper for the dispersion than for the third moment, in
agreement with theoretical predictions \citep{KS05,
  2010APh....32..340V}.

 Our smallest angular scale for second- (third-) order is $2 \, (5) \arcmin$.
This corresponds to a Fourier scale $\ell$ of about $2,000 \, (900)$. Since the
filter functions decay exponentially, we are not sensitive to $\ell >
10,000 \, (3,000)$. At the redshift of peak lensing efficiency ($z=0.4)$,
this corresponds to 3D Fourier scales of $k/[h/Mpc]$ of $6.6 \, (2)$. At
redshift of $z=0.1$, below which the large-scale structure contributes less
than 10\% to the overall lensing signal, the corresponding $k$-mode is $25 \,
(7) \, h$/Mpc.

\subsection{Calculating the 3PCF and binning of triangles}
\label{sec:binning}

\begin{figure}
  \centerline{Clone}

\resizebox{\hsize}{!}{
   \includegraphics{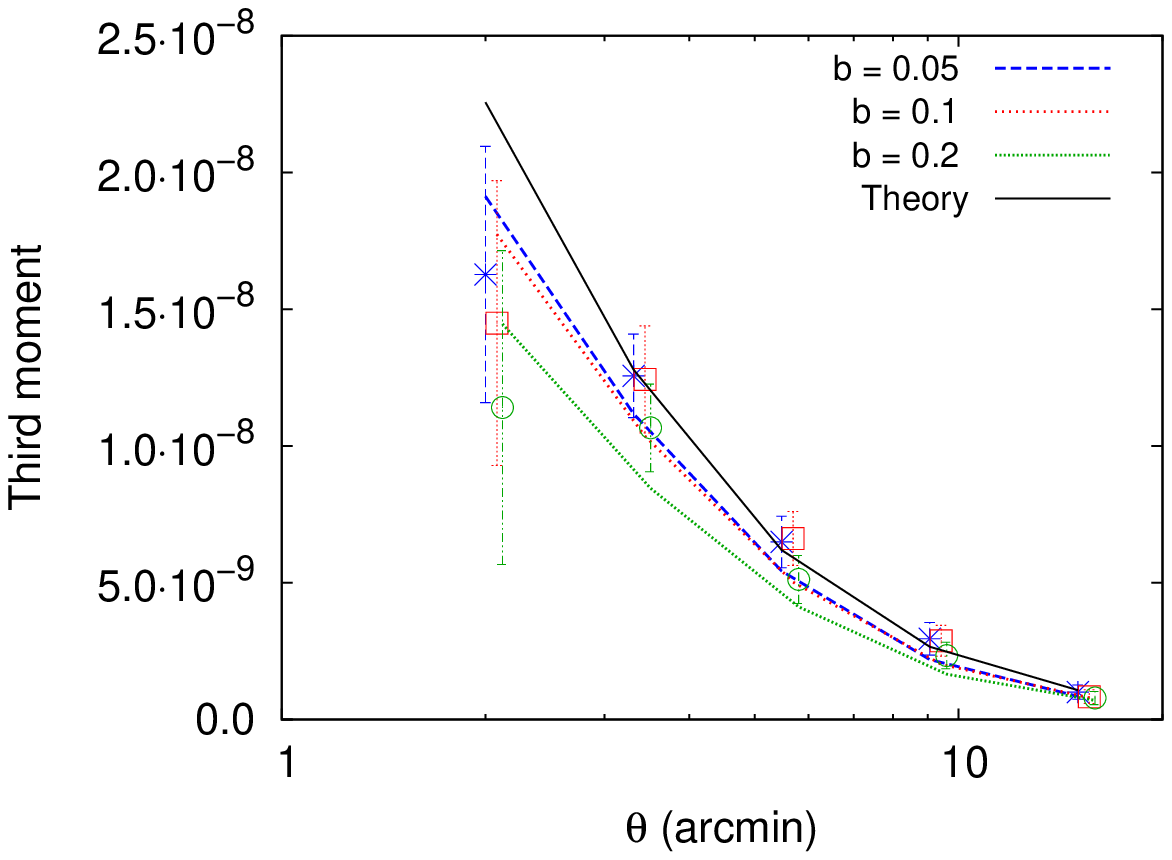}
}

 \centerline{CFHTLenS}

\resizebox{\hsize}{!}{
   \includegraphics{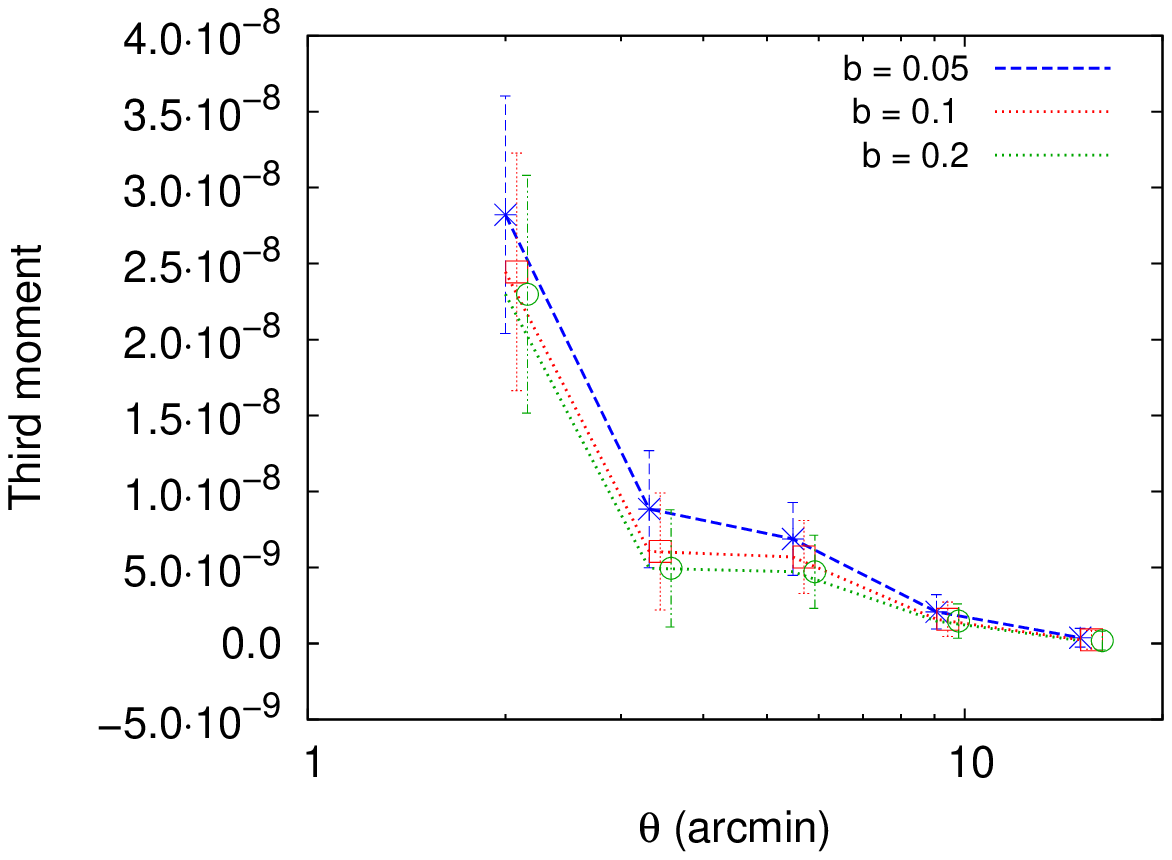}
}
\caption{\emph{Upper panel:} The third-order aperture-mass moment
  using 20 out of 184 simulated Clone fields with three different size
  of the binning of triangles: 0.05 (blue), 0.1 (red) and 0.2 (green),
  with (open symbols) and without (lines) noise. The theoretical
  prediction with the WMAP5 cosmology is shown as blue line. The error
  bars are the dispersion of 20 Clone simulations fields. \emph{Lower
    panel:} The aperture-mass skewness from CFHTLenS data. The lines
  for three different triangle bin sizes are the same as in the upper
  panel.  The error bars are calculated from the 184 independent Clone
  fields of view, rescaled to the observed survey area, and contain
  Poisson noise and cosmic variance.}
\label{fig:e3.clone} 
\end{figure}

We obtain the third-order aperture-mass moment from the shape
catalogue by integrating over the measured and binned three-point
shear correlation function (3PCF). The computation time, even using a
fast tree code, limits us in the use of very small bins.  The
tree-code from JBJ04 uses equidistant bins of size $b$ in the
logarithm of the triangle side lengths. We explore various bin sizes
$b = 0.2, 0.1, 0.05, 0.04$ using a subsample of 20 out of 184
simulated Clone fields. As shown in the top panel of
Fig~\ref{fig:e3.clone}, large bin sizes underestimate the
signal. However, we find convergence of the results for bin sizes
smaller than $b=0.1$.  Based on these findings, we choose the
conservative bin size $b=0.05$ for the CFHTLenS data.

When applying bin sizes $b=0.1$ and $0.05$ to CFHTLenS data, we do see
slightly larger differences in the resulting amplitude of $\langle
M_{\rm ap}^3 \rangle$ (see the bottom panel of
Fig~\ref{fig:e3.clone}). The differences are well within the
statistical uncertainty of the data, and we leave a more detailed
exploration of the 3PCF calculation for future work.

\subsection{Distribution of the third moment measurements}
\label{sec:distribution}

We use a Gaussian likelihood function (\ref{log_likeli}) of our data
as an approximation of the true likelihood function.
\cite{CFHTLenS-3pt} consider a non-Gaussian treatment of the
likelihood function using an independent component analysis (ICA) with
the help of numerical simulations. They found that the effect on
cosmological parameters is minor. Thus, the Gaussian likelihood
functions still represents a good approximation.

To further test this assumption, we compute the distribution of the
aperture-mass skewness $\langle M_{ap}^3\rangle(\theta)$ from $n =
184$ realisations of the Clone including intrinsic galaxy ellipticity
noise. The results are printed in Table \ref{tab:map3_moments}. The
rms of the skewness and kurtosis from a Gaussian distribution with
unknown mean are $\sqrt{6/n}$ and $\sqrt{24/n}$, respectively. There
is a marginal detection of a negative kurtosis at large scales. The
skewness is consistent with zero. We conclude that the assumption of a
Gaussian distribution for $\langle M_{ap}^3\rangle(\theta)$ is
sufficient for our purpose.

\begin{table}
  \caption{Skewness and kurtosis of $\langle M_{ap}^3 \rangle(\theta)$
    for three different smoothing scales, measured from 184 lines of
    sight of CFHTLenS clone simulations.  The errors assume a Gaussian
    distribution.}

\begin{center}
\begin{tabular}{@{}lr@{ $\pm$ }lr@{ $\pm$ }lr@{ $\pm $ }l} \hline
Scale $\theta$          & \multicolumn{2}{@{}c}{$1\arcmin$} & \multicolumn{2}{@{}c}{$10\arcmin$} & 
\multicolumn{2}{@{}c}{$20\arcmin$} \\ \hline
Skewness                & $0.68$  & $0.55$  & $0.12$ & $0.55$   & $-0.058$ & $0.55$ \\
Kurtosis                & $-0.14$ & $1.1$   & $-1.4$ & $1.1$    & $-1.4$   & $1.1$\\
\hline\\

\end{tabular}
\label{tab:map3_moments}
\end{center}

\end{table}

\end{appendix}

\end{document}